\documentclass[iop]{emulateapj}
\usepackage{graphicx, lmodern, dcolumn}
\usepackage[T1]{fontenc}
\usepackage[para,flushleft]{threeparttable}
\usepackage{amssymb,amsmath}
\newcommand{\lapprox }{{\lower0.8ex\hbox{$\buildrel <\over\sim$}}}
\newcommand{\gapprox }{{\lower0.8ex\hbox{$\buildrel >\over\sim$}}}

\shorttitle{Stellar Feedback in Galaxy Simulations}
\shortauthors{A. N\'u\~nez et al.}

\begin{document}
\title{Modeling for Stellar Feedback in Galaxy Formation Simulations}
\author{Alejandro N\'u\~nez\altaffilmark{1}, Jeremiah P. Ostriker\altaffilmark{1,2}, Thorsten Naab\altaffilmark{3}, Ludwig Oser\altaffilmark{3}, Chia-Yu Hu\altaffilmark{3} and Ena Choi\altaffilmark{4}}

\altaffiltext{1}{Department of Astronomy, Columbia University, 550 West 120th Street, New York, NY 10027, USA}
\altaffiltext{2}{Princeton University Observatory, Princeton, NJ 08544, USA}
\altaffiltext{3}{Max-Planck-Institut f\"ur Astrophysik, Karl-Schwarzschild-Strasse 1, D-85741 Garching, Germany}
\altaffiltext{4}{Department of Physics and Astronomy, Rutgers, The State University of New Jersey, Piscataway, NY 08854, USA}

\begin{abstract}
Various heuristic approaches to model unresolved supernova (SN) feedback in galaxy formation simulations exist to reproduce the formation of spiral galaxies and the overall inefficient conversion of gas into stars. Some models, however, require resolution dependent scalings. We present a sub-resolution model representing the three major phases of supernova blast wave evolution ---free expansion, energy conserving Sedov-Taylor, and momentum conserving snowplow--- with energy scalings adopted from high-resolution interstellar-medium simulations in both uniform and multiphase media. We allow for the effects of significantly enhanced SN remnant propagation in a multiphase medium with the cooling radius scaling with the hot volume fraction, $f_{\mathrm{hot}}$, as $(1 - f_{\mathrm{hot}})^{-4/5}$. We also include winds from young massive stars and AGB stars, Str\"omgren sphere gas heating by massive stars, and a gas cooling limiting mechanism driven by radiative recombination of dense H\textsc{ii} regions. We present initial tests for isolated Milky-Way like systems simulated with the \textsc{Gadget} based code SPHgal with improved SPH prescription. Compared to pure thermal SN input, the model significantly suppresses star formation at early epochs, with star formation extended both in time and space in better accord with observations. Compared to models with pure thermal SN feedback, the age at which half the stellar mass is assembled increases by a factor of 2.4, and the mass loading parameter and gas outflow rate from the galactic disk increase by a factor of 2. Simulation results are converged for a two order of magnitude variation in particle mass in the range (1.3--130)$\times 10^4$ solar masses. 
\end{abstract}

\keywords{methods: numerical -- galaxies: formation -- galaxies: evolution -- galaxies: stellar content -- galaxies: abundances}

\section{Introduction}
Current cosmological simulations of baryonic and dark matter (DM) have made great advances in recreating the epochs and spatial distribution of galaxies \citep[see][for reviews]{Bertschinger98, Springel10, Somerville15}. A major challenge is the reproduction of the overall inefficiency of galaxy formation with physically self-consistent models. Typically, there is a tendency for gas in simulated galaxies to transform into stars too efficiently and too early, resulting in stellar mass fractions larger than observed and stellar populations older than observed \citep[e.g.,][]{White91, Keres09, Guo10, Moster13}.

In the case of low mass galaxies, early work has pointed at efficient feedback from supernova (SN) explosions as the appropriate solution, as it would suppress stellar formation at early epochs by heating up and diffusing gas, resulting in systems with lower stellar mass fractions and younger stellar ages \citep{White91}. The most widely used feedback is SN-driven outflows \citep{Dekel86, Larson74}, and such flows are detected in many observations at low and high redshifts \citep[e.g.,][]{Martin99, Strickland00, Rupke05, Steidel10, Newman12}. Considering the large variations in numerical implementations to create galactic outflows \citep[e.g.,][]{Springel03, Dubois08, Hopkins14, Dave13, Teyssier13, Vogelsberger14, Kimm2015, Schaye15, Simpson15} in galactic and cosmological simulations, it appears that it is not yet clear how best to implement SN feedback into simulations with limited resolution, which cannot fully resolve individual SN explosions and turbulent interstellar medium (ISM).

The simplest implementation of SN feedback is to inject thermal energy locally to the surrounding medium. This energy, however, is quickly radiated away in dense regions and the blast wave evolution is followed incorrectly \citep{Navarro93} \citep[see][for attempts to follow the blast wave evolution by turning off ``feedback'' or ``radiative losses'']{Stinson06}. This situation is an artifact of low resolution simulations, preventing previous SN explosions from vacating their surroundings, thus keeping subsequent SN explosions in dense areas \citep[see e.g.,][]{Dalla12, Kimm2015}. Some studies attempt to avoid this by applying a stochastic stellar feedback that is, first, calibrated to match the observed galaxy stellar mass function and, second, distributed over less mass than heating all neighbors, which results in longer cooling times \citep[see studies with the Evolution and Assembly of Galaxies and their Environments ---\textsc{Eagle}--- simulation, e.g.,][]{Schaye15, Furlong2015}.

A second order approach consists of transferring part or all of SN energy as kinetic energy to the surrounding medium by conserving momentum from the initial explosion \citep[see e.g.,][]{Dubois08b}. In this scenario, however, most of the energy transferred might still end up in thermal form, since the ratio of transferred kinetic energy to total SN energy is proportional to the ratio of transferred SN ejecta mass to the mass of the surrounding medium when momentum conservation is imposed \citep[see e.g.,][]{Hu14, Kimm2015}. Thus, while this approach explicitly includes the momentum of the ejecta, it effectively evolves to the usual ``thermal feedback'' algorithm. Therefore, overcooling still plays a part in hindering SN energy feedback from noticeably affecting the structure and evolution of simulated galaxies.

To prevent overcooling from nullifying SN energy feedback, recent studies adopted approximate approaches, such as decoupling SN-affected gas from hydrodynamic interactions \citep[e.g.,][]{Springel05a, Oppenheimer08, Vogelsberger13, Vogelsberger14}, implementing decoupled multi-phase ISM \citep{Scannapieco06, Aumer13}, suppressing radiative cooling effects on SN-heated gas for several tens of Myr \citep{Stinson06, Guedes2011}, constructing feedback mechanisms based on the evolution of superbubbles \citep{Keller2014, Keller2015}, and stochastically enhancing heating of SN-affected gas \citep{Dalla12, Schaye15}. As encouraging as the results are at replicating SN-driven winds and star formation rates (SFR), the methods have limited predictive power as they are not based on \emph{ab initio} physical modeling of the stellar feedback process.

In this study we attempt to develop a numerical stellar feedback prescription based as closely as possible on standard physical principles. It includes winds and heating from young massive stars as well as feedback from SN explosions, the latter following the known physical processes that govern the different phases of SN remnants. We use the smoothed particle hydrodynamics (SPH) code SPHGal \citep{Hu14}, an implementation that incorporates several improvements into the \textsc{Gadget} code \citep{Springel05b} to address some known SPH problems. Our approach is most similar to the stellar feedback in \citet{Agertz13} and \citet{Agertz2016}, as well as the Feedback in Realistic Environment (\textsc{Fire}) project, presented in \citet{Hopkins14} and used by \citet{Hopkins12, Hopkins2013b}, \citet{Muratov15}, and \citet{Su2016}. We apply our stellar feedback to an isolated Milky Way-like galaxy within an isolated dark halo. The tools developed here are in use by the cosmological code in \citet{Choi2016}.

We break our approach into three parts: feedback from young, pre-SN stars, feedback from dying low-mass stars, and feedback from SN events. (1) We include the effects of stellar winds from young stars by transferring momentum to neighboring gas particles for the first few Myr. We also heat gas particles within the Str\"omgren sphere of young star particles, mimicking the effects of ionizing radiation from massive stars \citep[see e.g.,][]{Renaud13}. (2) We transfer momentum from old, low-mass stars to the surrounding gas throughout several Gyr after a type II SN has occurred, to imitate the slow winds from asymptotic giant branch (AGB) stars.

(3) In our SN feedback model, we transfer SN energy to the closest gas particles to the SN event, each one receiving mass and energy according to a computed high resolution evolution of SN remnants (SNR). Detailed, very high resolution hydrodynamic simulations of SNR evolution \citep[see][for details and a review]{Li15} show that they pass through three successive phases: free expansion (FE), energy-conserving Sedov-Taylor (ST), and momentum-conserving snowplow (SP) \citep[see also][]{Kim15, Martizzi15, Walch15}. In our model, which we call the SP-model, each neighboring gas particle receives energy following the physics governing the SNR phase in which the gas particles lie. If the gas particle lies within the radius of the ejecta-dominated FE phase, SN energy is transferred by conserving the ejecta momentum. If the gas particle lies outside the FE radius but within the radius of the adiabatic blast wave ST phase, SN energy is transferred 30\% as kinetic energy and 70\% as thermal energy. Finally, if the gas particle is beyond the ST radius and into the realm of the pressure driven and momentum conserving SP phase \citep{Chevalier74}, then part of the original SN energy is dissipated and only a fraction is transferred to the gas particle, with the thermal portion dissipating more quickly than the kinetic portion following a prescription based on the detailed, high resolution, hydrodynamic treatments of SNR evolution in \citet{Li15}. Our treatment of feedback in the SP phase allows for a multi-phase medium by allowing for the volume fraction that is hot and tenuous.

The paper is organized as follows: in Sec.~\ref{sec:sim} we describe our numerical code simulation setup and feedback implementation, in Sec.~\ref{sec:results} we present our results, and in Sec.~\ref{sec:conclusions} we discuss these results and possible further developments.

\section{Simulations}\label{sec:sim}
\subsection{Numerical Code}\label{subsec:code}
To test our SN feedback treatment, we used a recently modified version of \textsc{Gadget} \citep{Springel05b}. The improved version ---SPHGal--- is described in detail by \citet{Hu14}. SPHGal includes an artificial viscosity implementation modified from the version presented in \citet{Cullen10} that uses a shock indicator based on the time derivative of the velocity divergence so as to improve the shock treatment. It also includes an energy diffusion implementation based on \citet{Read12} to correct for mixing instabilities. Finally, it includes a switch from a density--entropy to a pressure--entropy formulation to more accurately compute contact discontinuities as described in \citet{Hopkins2013a}, \citet{Ritchie01}, and \citet{Saitoh13}. As in the original \textsc{Gadget} code, SPHGal calculates the gas dynamics using the Lagrangian SPH technique \citep[see e.g.,][]{Monaghan92} and it ensures the conservation of energy and entropy \citep{Springel02}.

To complement the original \textsc{Gadget}, in which star formation and cooling was computed for a primordial composition of hydrogen and helium \citep{Katz96}, we included metal-line cooling using the method and cooling rates described in \citet{Aumer13}, where net cooling rates are tracked for 11 species: H, He, C, Ca, O, N, Ne, Mg, S, Si, and Fe. This method exposes the gas to the redshift-dependent UV/X-ray and cosmic microwave background modeled in \citet{Haardt01} and allows for mixing and spreading of metal abundances in the enriched gas. The code assumes the gas to begin with zero metallicity and to be optically thin and in ionization equilibrium.

\subsection{Star Formation Model}\label{subsec:sfr}
We included a chemical evolution and stochastic star formation model \citep{Aumer13} that considers enrichment by type II SNe, type Ia SNe, and AGB-type winds following the chemical yield prescriptions of \citet{Woosley95}, \citet{Karakas10}, and \citet{Iwamoto99}, respectively (see Secs.~\ref{subsubsec:implementation}--\ref{subsec:AGB}). As described in \citet{Aumer13}, SFR is set by $d\rho_*/dt = \varepsilon \rho\mathrm{_{gas}}/t\mathrm{_{dyn}}$, where $\rho_*$ and $\rho\mathrm{_{gas}}$ are the stellar and gas densities, respectively, $\varepsilon$ is the star formation efficiency, and $t\mathrm{_{dyn}}$ = $1/\sqrt{4\pi G\rho\mathrm{_{gas}}}$ is the local dynamical time for the gas particle. This is applied to gas particles in a convergent flow that are denser than a density threshold $\rho_{\mathrm{th}}$, which we define as
\begin{equation}\label{rhothresh}
  \rho_{\mathrm{th}} = \rho_0 \left(\frac{T_{\mathrm{gas}}}{T_0}\right)^3 \left(\frac{M_0}{M_{\mathrm{gas}}}\right)^{2},
\end{equation}
where $\rho_0$ is a critical threshold density value, $T_0$ is a critical temperature threshold, $M_0$ is a low-resolution particle mass fiducial value, and $T_{\mathrm{gas}}$ and $M_{\mathrm{gas}}$ are the temperature and mass of the gas particles, respectively. The values of $\rho_0$, $T_0$, and $M_0$ used in our simulations are listed in Table~\ref{table:params}. This scaling is modeled on the requirement that the density must exceed the value for the Jeans gravitational instability of a mass $M_{\mathrm{gas}}$ at temperature $T_{\mathrm{gas}}$. Star particles are then created stochastically with one gas particle being turned into one star particle of the same mass.

\subsection{Radiative Recombination}\label{subsec:recombination}
To account for radiative recombination processes in the cool ($\lapprox$10$^4$ K) ISM, where H and He line cooling dominate, we implemented a simple cooling limiting mechanism based on the recombination time of dense H\textsc{ii} regions. The recombination time for a gas particle is defined by $t_{\mathrm{rec}}^{-1} = n_{\mathrm{H}}\ \alpha_{\mathrm{B}}$, where $n_{\mathrm{H}}$ is the hydrogen number density of the ISM in cm$^{-3}$ and $\alpha_{\mathrm{B}}\approx2.56\times 10^{-13}$ cm$^3$ s$^{-1}$ is the effective radiative recombination rate for hydrogen, assuming a sphere temperature of $10^4$ K \citep{Draine11}.

We calculate the gas temperature resulting from the cooling limiting effect at every time step using the following definition
\begin{equation}\label{Trec}
  T_{n+1,\mathrm{rec}} = T_n + \frac{\Delta t}{\Delta t + t_{\mathrm{rec}}}(T_{n+1} - T_n),
\end{equation}
where $T_n$ and $T_{n+1}$ are the gas particle temperatures as determined by the cooling rates described in \citet{Aumer13} (see Sec.~\ref{subsec:code}) for the previous and current time steps $n$ and $n+1$, respectively, and $\Delta t$ is the time step size. If the resulting recombination-limited temperature $T_{n+1,\mathrm{rec}}$ is below $10^4$ K, then the gas particle gets assigned $T_{n+1,\mathrm{rec}}$ instead of $T_{n+1}$. The purpose of this limitation on the cooling rate is to allow for the fact that the relevant atomic processes (i.e., H and He line cooling) cannot proceed faster than the recombination time even if the code assumes (for simplicity) equilibrium states of ionization.

\subsection{Early Stellar Feedback}\label{subsec:youngfb}
As described by several stellar evolution models \citep[e.g.,][]{Leitherer99}, almost as much momentum is emitted into the ISM by winds from young massive stars as by type II SN explosions. Furthermore, radiation from these massive stars heats and ionizes the surrounding gas. The effect of this ionizing radiation can be characterized by the Str\"omgren sphere \citep{Stromgren39} surrounding massive stars, within which the ISM is ionized and heated to $\approx$10$^4$ K \citep{Hopkins12, Renaud13}.

\citet{Stinson13} presented an ``early stellar feedback'' model that adds thermal energy to gas surrounding stars, but only considering the energy in the UV stellar flux ---approximately 10\% of the total stellar luminosity \citep{Leitherer99}. To have a more complete model for pre-SN stellar feedback, we included two distinct feedback mechanisms from young massive stars before they explode as SN: stellar winds and Str\"omgren sphere heating \citep[see e.g.,][]{Agertz13, Renaud13}.

We implemented a simple momentum conservation energy transfer of winds from massive stars, to first approximation using the same amount of ejected mass and ejecta velocity as those of type II SN explosions (see Sec.~\ref{subsubsec:implementation}) evenly spread in time from star birth up to the moment of SN explosion $t_{\mathrm{SN}}$ \citep[for a detailed description of stellar winds see][]{Kudritzki00}.

We also included Str\"omgren sphere heating by adjusting the temperature of cold gas particles within the Str\"omgren radius of a star particle, defined as
\begin{equation}\label{stromgren}
  R_{\mathrm{Str}} = \left(\frac{3}{4\pi}\frac{\dot Q}{{n_{\mathrm{H}}^2} \alpha_{\mathrm{B}}}\right)^{1/3},
\end{equation}
where $\dot Q$ is the ionizing photon flux and equals $10^{49} \Delta N$, with $\Delta N$ being the number of stars to explode as type II SN. To first approximation, the temperature of cold gas particles (i.e., with $\lapprox10^4$ K) within the Str\"omgren sphere of a star particle is adjusted such that
\begin{equation}\label{Tstromgren}
  T_{\mathrm{new}} = T_{\mathrm{old}} + (10^4 - T_{\mathrm{old}}) \frac{\Delta t}{t_{\mathrm{ion}}},
\end{equation}
where $T_{\mathrm{new}}$ and $T_{\mathrm{old}}$ are the updated and original gas particle temperatures in Kelvin, respectively. Assuming ionization equilibrium, we set $t_{\mathrm{ion}}=t_{\mathrm{rec}}$.

\subsection{SN Energy Feedback}\label{subsec:snfeedback}

\subsubsection{SNR Phases in Detail}\label{subsubsec:phases}
Our motivation for developing a snowplow phase driven SN feedback is based on the current spatial and temporal resolutions of simulations, which are far from being able to resolve individual SN events and the energy transfer that occurs in their immediate vicinity in time and space \citep[for a general review of the SNR phases, see][]{Ostriker88}. The typical duration of the first phase of a SNR ---the FE phase--- is limited to a time $t_{ST}\approx200$ yr, beyond which the Sedov-Taylor phase begins  \citep[see equation 39.3 of][]{Draine11}. During time $t$ \textless\ $t_{\mathrm{ST}}$ the ejected mass is much greater than the swept up mass and to lowest order it expands at nearly constant velocity, conserving momentum as it hits the surrounding medium.

The radius at which the FE phase ends lies at the point where the total mass swept by the SNR blast wave equals the mass of the SNR itself. This occurs at a radius
\begin{equation}\label{FEradius}
  r_{\mathrm{ST}} = \left(\frac{3}{4}\ \frac{M_{\mathrm{ej}}}{\pi\rho}\right)^{1/3},
\end{equation}
where $M_{\mathrm{ej}}$ is the mass ejected in the SN event and $\rho$ is the mass density of the surrounding medium. SPH simulations with particles masses $\approx$10$^5$ M$_{\sun}$ numerically result in $r_{\mathrm{ST}}$ in the order of tens of parsecs. Clearly, implementing SN feedback based on pure momentum conservation of the SN ejecta, as it applies to the FE phase, would be hitting the limits of spatial resolution, which is typically larger than 100 pc.

During the ST phase the blast wave expands adiabatically, experiences negligible radiative loses, and is described by a self-similar solution \citep{Taylor50, Sedov59, Bandiera84}. The kinetic energy of the expanding material is fixed at $\approx$30\% of the total energy. Radiative cooling becomes significant when the blast wave expands beyond the radius
\begin{equation}\label{STradius}
  r_{\mathrm{cool}} \simeq (19.1\ \textrm{pc})\ E_{51}^{5/17}\ n_0^{-7/17}\ (1-f_{\mathrm{hot}})^{-4/5},
\end{equation}
where $E_{51}$ is the kinetic energy of the ejecta normalized to 10$^{51}$ erg, $n_0$ is the number density of the surrounding (primarily) hydrogen medium\footnote{\citet{Blondin98} tested SN remnant propagation in the $n_0$ range 0.084--8400 cm$^{-3}$. They found unrealistic results only for the highest-density case, which was driven by their assumption of no magnetic fields.} in cm$^{-3}$, as described in \citet{Blondin98}, and $f_{\mathrm{hot}}$ is the volume fraction of the surrounding medium that is hot and has a very low density. The dependence of $r_{\mathrm{cool}}$ on $(1-f_{\mathrm{hot}})^{-4/5}$ is obtained from analysis of the high resolution simulations of \citet{Li15}. That work also found that in self-consistently generated multiphase media there was a strong correlation between the volume averaged temperature $\langle T\rangle$ and the filling factor of the hot gas, which is well approximated\footnote{This approximation is valid mostly for gas with $T \gapprox 2\times 10^5$ K.} by
\begin{equation}\label{fhot}
  f_{\mathrm{hot}} = 1 - exp[- (\langle T\rangle / T_f)^{2/3}],
\end{equation}
where $T_f=2\times10^6$ K. With this formulation, $f_{\mathrm{hot}}$ tends to zero in cold regions and to near unity in hot regions. Correspondingly, $r_{\mathrm{cool}}\propto exp(\langle T\rangle^{8/15})$, and so this transition radius increases with increasing temperature. SPH simulations with particles masses $\approx$10$^5$ M$_{\sun}$ numerically result in $r_{\mathrm{cool}}$ values at least $\sim$3 $r_{\mathrm{ST}}$ but can increase by orders of magnitude for low $n_0$ or large $f_{\mathrm{hot}}$ values.

Beyond $r_{\mathrm{cool}}$ the SNR enters the SP phase, in which the blast wave experiences a significant energy loss via radiative cooling and expands with relatively constant radial momentum \citep{Cox72, Cioffi88, Petruk06}. Residual thermal energy within the blast wave allows the momentum to increase slightly in this phase and, according to \citet{Blondin98}, is proportional to $t^{1/3}$. Based primarily on the detailed hydrodynamic simulations of \citet{Li15}, we describe the thermal energy to transfer from the SNR to the surrounding medium as
\begin{equation}\label{E_Th}
  E_{\mathrm{Th}} = \frac{0.7 E_{\mathrm{SN}}}{1 + 0.5(r / r_{\mathrm{cool}})^6},
\end{equation}
where $E_{\mathrm{SN}}$ is the total original kinetic energy of the SNR and $r$ is the distance in pc to the medium. The 0.7 factor accounts for the fact that, at the start of the SP phase, the thermal fraction of the total energy is $\approx$70\%.

Similarly, we describe the kinetic energy to transfer from the SNR to the surrounding medium as
\begin{equation}\label{E_K}
  E_{\mathrm{K}} = \frac{0.3 E_{\mathrm{SN}}}{1 + 0.18(r / r_{\mathrm{cool}})^3}.
\end{equation}
Note that since the exponent of the radius fraction is smaller in Eq.~\ref{E_K} than in Eq.~\ref{E_Th}, the late stages of the SP phase are dominated by its kinetic energy.

The resulting deceleration parameter is $vt/r\approx$ 0.28 in the type II SN scenario and $\approx$ 0.47 in the type I SN scenario, where $v$ is the velocity of the blast wave. The values that we have adopted are consistent with those recently found by \citet{Kim15} \citep[see also][]{Walch15, Martizzi15}, who also found a more rapid deceleration and energy loss than the value of 0.33 found by \cite{Blondin98} \citep[see also][for ISM simulations accounting for the effect of feedback during the SP phase]{Walch15, Girichidis16}.

\subsubsection{Construction of a New SN Feedback Model}\label{subsubsec:implementation}
We construct a feedback formulation that considers the three different SNR phases described in Sec.~\ref{subsubsec:phases}, and we call it \textbf{\textit{SP-model}} feedback, since it includes the SP phase of SNR evolution. This model can be applied to both type II and Ia SN explosions. We allow the model to transfer SN ejecta energy to the closest 10 neighboring gas particles according to the physics of the SNR phase in which the neighboring particles happens to lie according to the following recipe:
\begin{itemize}
\item If the distance between the gas particle and the SN event is smaller than $r_{\mathrm{ST}}$ (Eq.~\ref{FEradius}), SN energy is transferred by conserving momentum, which results in a small fraction of the original SN energy being transferred as kinetic energy and the rest as thermal energy; we call this \textit{FE feedback}.
\item If the distance lies between $r_{\mathrm{ST}}$ and $r_{\mathrm{cool}}$ (Eq.~\ref{STradius}), 30\% of the SN energy is transferred as kinetic energy and the rest as thermal energy; we call this \textit{ST feedback}.
\item If the distance is larger than $r_{\mathrm{cool}}$, SN energy is transferred as both kinetic and thermal energy following Eqs. \ref{E_Th} and \ref{E_K}; we call this \textit{SP feedback}.
\end{itemize}

The total energy liberated in a typical Type II SN is on the order of 10$^{53}$ erg, but much of the loss is emitted as neutrinos and only 1\% goes out as kinetic energy with the ejecta, or $\sim$10$^{51}$ erg. This canonical value for SN energy in the ejected material is described as
\begin{equation}\label{E_SN}
  E_{\mathrm{SN}} = \frac{1}{2}\ M_{\mathrm{ej}}\ v_{\mathrm{ej}}^2,
\end{equation}
where $v_{\mathrm{ej}}$ is the velocity of the ejecta \citep{Janka07}. In our simulations we calculate $M_{\mathrm{ej}}$ following the mass yield prescriptions for type II SNe by \citet{Woosley95} and for type Ia SNe by \citet{Iwamoto99}, the former resulting in a typical ejected mass of $\approx$13\% of the exploding star. Since only stars with masses between 8 and 100 M$_{\sun}$ experience type II SNe, the expected value for $v_{\mathrm{ej}}$ falls in the range 3000--10000 km s$^{-1}$. We choose $v_{\mathrm{ej}}=4500$ km s$^{-1}$ as our fiducial value for both type Ia and II SNe.

Note that each SN event in our simulation corresponds to the energy of many simultaneous physical SN events, as mandated by simulation resolution \citep[see, however,][for dwarf galaxy simulations resolving individual SNe]{Hu15}. For a star particle of mass $\sim$10$^5$ M$_{\sun}$, the typical value of $E_{\mathrm{SN}}$ is in the order of 10$^{55}$ erg. Fortunately, the same way that Eq. \ref{E_SN} scales with particle mass, so do Eqs. \ref{FEradius}--\ref{E_K}, making the SP-model formulation formally independent of simulation resolution.

\subsection{Winds From AGB Stars}\label{subsec:AGB}
As part of the SP-model feedback, we also include feedback from the low-mass end of the stellar population in the form of slow winds from AGB stars. Similar to our treatment of winds and metals from young massive stars (see Sec.~\ref{subsec:youngfb}), we transfer energy from the star particles to the surrounding gas particles by conserving momentum of the ejected material, which is calculated using the mass yield prescriptions for AGB by \citet{Karakas10}. We assume an AGB wind velocity of $v_{\mathrm{AGB}}=10$ km s$^{-1}$.

\subsection{Isolated Galaxy Simulation}\label{subsec:isolated}
With our simulated galaxy we are recreating a Milky Way-like spiral galaxy isolated from any outside interference to isolate the effects of stellar feedback on the evolution of the galaxy dynamics and structure. The galaxy consists of a DM halo, a rotationally supported disk of gas and stars, a central stellar bulge, and a hot gas halo. We followed the method described in \citet{Springel05a} and extended by \citet{Moster11} for a hot gas halo to set up close to equilibrium initial conditions.

There are $3\times 10^5$ DM particles with mass $4.5\times 10^6$ M$_{\sun}$, for a total DM mass of $1.3\times 10^{12}$ M$_{\sun}$ and they are arranged so as to follow the mass distribution profile described by \citet{Hernquist90} with a concentration parameter $c=9$ \citep{Navarro97}.

The total baryonic mass is $5.6\times 10^{10}$ M$_{\sun}$ and is represented by three different galactic structures: a disk of gas and stars, a stellar bulge, and a hot ($\gapprox10^6$ K) gas halo. Almost 80\% of the disk is in a stellar disk, while the other 20\% is represented by $9.5\times 10^4$ collisional SPH gas particles each of mass $1.3\times 10^5$ M$_{\sun}$. The total mass of the stellar bulge is $1.0\times 10^{10}$ M$_{\sun}$. The total mass of the hot gas halo is $2\times 10^{9}$ M$_{\sun}$, approximately 4\% of the total baryonic mass of the galaxy.

We include this diffuse, rotating, hot halo ---as described in \citet{Moster11} \citep[see also][]{Choi14}--- to recreate observed, X-ray emitting, extended hot haloes around moderate mass, spiral galaxies \citep{Anderson13}. The gas halo assumes that there is no angular momentum transport between it and the spherical DM halo. The median temperature of the hot halo gas is $1.3\times 10^6$ K and the mean hydrogen density is $6.3\times 10^{-5}$ cm$^{-3}$, similar to the values found for the Milky Way Galaxy by \citet{Faerman2016}.

The total galactic mass is 1.4 x $10^{12}$ M$_{\sun}$. Up to 95\% of the total galactic mass is contained within 1 Mpc from the galactic center, and 100\%, within 8 Mpc. The side length of the cube box is 50 Mpc. Table~\ref{table:IC_isolated} summarizes SPH particle characteristics for our fiducial initial conditions.

\begin{table}
\centering
\caption{SPH Particles Description of Initial Condition for our Fiducial Milky-Way Like Isolated Galaxy.}
\label{table:IC_isolated}
\begin{threeparttable}
\newcolumntype{d}{D{.}{.}{-1}}
\begin{tabular}{@{}rddd@{}d}
\hline\hline
 & & \multicolumn{1}{c}{Mass} & \multicolumn{1}{c}{Mass} \\
\multicolumn{1}{c}{Particle} & \multicolumn{1}{c}{Number} & \multicolumn{1}{c}{per\ particle} & \multicolumn{1}{c}{Total} & \multicolumn{1}{c}{$\epsilon$ \tnote{a}} \\
\multicolumn{1}{c}{type} & \multicolumn{1}{c}{(10$^4$)} & \multicolumn{1}{c}{(10$^5$ M$_{\sun}$)} & \multicolumn{1}{c}{(10$^{10}$ M$_{\sun}$)} & \multicolumn{1}{c}{(kpc)} \\[0.05 in]
\hline
\\[-1.5mm]
Halo gas    & 1.7  & 1.3  & 0.2   & 0.044 \\
Disk gas    & 9.5  & 1.3  & 1.2   & 0.044 \\
Disk stars  & 24.0 & 1.3  & 3.1   & 0.044 \\
Bulge stars & 7.5  & 1.3  & 1.0   & 0.044 \\
DM          & 30.0 & 44.7 & 134.0 & 0.220 \\[0.05 in]
\hline
\\[-1.5mm]
Baryonic & & & 5.6 \\
Total & & & 139.5 \\
\hline\hline
\end{tabular}
\begin{tablenotes}
\item[a] Gravitational softening length.
\end{tablenotes}
\end{threeparttable}
\end{table}

The galaxy is evolved for 6 Gyr using the fixed gravitational softening lengths specified in Table~\ref{table:IC_isolated} for each particle type. The gas is converted to stars according to the prescription in Sec.~\ref{subsec:sfr} and using the fiducial parameter values described in Table~\ref{table:params}. The same softening length that applies to gas particles also applies to new star particles (see Table~\ref{table:IC_isolated}). Star particles then stochastically experience one type II SN event after an age $t_{\mathrm{SN}}=3$ Myr. They also experience continuous type Ia SN events and AGB feedback starting at age 50 Myr and every 50 Myr after that (see Secs.~\ref{subsubsec:implementation}--\ref{subsec:AGB}). The disk and bulge stars from the initial conditions do not contribute to any feedback at any point in the simulation. All input parameters for the simulation are summarized in Table~\ref{table:params}.

\begin{table*}
\centering
\caption{Parameters Used in Simulation.}
\label{table:params}
\begin{threeparttable}
\begin{tabular}{c@{\ \ }c@{\ \ }l@{\ \ \ \ \ }}
\hline\hline
Symbol & Fiducial Value & Note \\
\hline
\\[-1.5mm]
\multicolumn{1}{l}{\it Star Formation} \\
$\varepsilon$ & 0.025 & Star formation efficiency \\
$\rho_0$ & 1 cm$^{-3}$ & Critical density threshold \\
$T_0$ & 6000 K & Critical temperature threshold \\
$M_0$ & 1.3$\times 10^6$ M$_{\sun}$ & Low resolution particle mass fiducial value \\
 \\
\multicolumn{3}{l}{\it Radiative Recombination and Early Stellar Feedback} \\
$\alpha_{\mathrm{B}}$ & 2.56$\times10^{-13}$ cm$^3$ s$^{-1}$ & Effective radiative recombination rate for hydrogen\\
$t_{\mathrm{SN}}$ & $3\times 10^6$ yr & Lifetime of massive stars before they explode as type II SN \\
 \\
\multicolumn{1}{l}{\it SN and AGB Feedback} \\
$v_{\mathrm{ej}}$ & 4500 km s$^{-1}$ & Velocity of ejected SN mass and stellar wind mass \\
$v_{\mathrm{AGB}}$ & 10 km s$^{-1}$ & Velocity of ejected AGB mass \\
$T_f$ &  $2\times10^6$ K & Temperature factor for the calculation of $f_{\mathrm{hot}}$. \\[0.05 in]
\hline\hline
\end{tabular}
\end{threeparttable}
\end{table*}

Lastly, mass and energy emanating from a star particle from one of the events described in Sec~\ref{subsec:youngfb}--\ref{subsec:AGB} is distributed among neighboring gas particles according to their distance from the star particle using weights calculated by the Wendland $C^4$ kernel \citep{Dehnen12}. All velocities added to neighbor gas particles point radially away from the location of the star particle. This is true for stellar winds from O and B stars, Str\"omgren sphere heating, type II and Ia SNe, and AGB winds. We find that at the end of our simulations the mass range of the newly formed star particles is 0.7--13.4$\times 10^5$ M$_{\sun}$, with a median value of 0.9 $\times 10^5$ M$_{\sun}$. The details of the implementation of the SP-model to SPH simulations are included in Appendix~\ref{appendix}.

\section{Results}\label{sec:results}
In addition to a simulation using the SP-model feedback formulation, we ran two other simulations with the same initial conditions but different feedback schemes: \textbf{\textit{NO feedback}}, in which we apply mass transfer but no energy transfer at all, and \textbf{\textit{Hot bubble}} feedback, in which we transfer SN energy in purely thermal form to the five closest neighboring particles. Note that, by construct, the amount of total kinetic energy transferred is null in these two alternative scenarios. We compare the results from these two alternative scenarios to the SP-model results to better evaluate the effect of the SP-model. The most relevant results from our isolated galaxy simulations are summarized in Table~\ref{table:results}.

\begin{figure}
\includegraphics{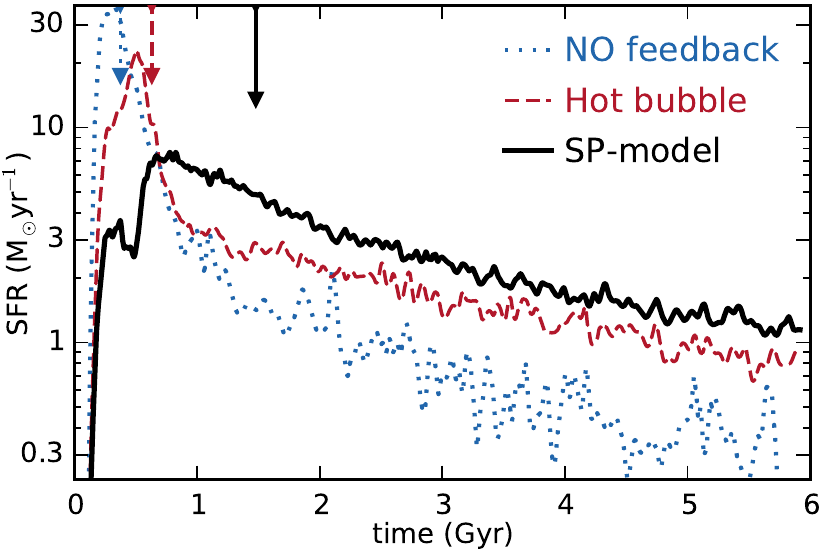}
\caption{SFR as a function of time for our Milky Way-like isolated galaxy using three different SN feedback schemes: no energy feedback (NO feedback), purely thermal feedback to the five closest neighboring particles (Hot bubble), and the model SN energy feedback (SP-model), which transfers energy according to the physics of the SN phase (free expansion, Sedov-Taylor, or snowplow) in which each of the closest 10 neighboring particles lies. The vertical arrows indicate the age at which half the new galactic stellar mass has been assembled.}
\label{fig:SFR_isolated}
\end{figure}

\subsection{SFR and Galactic Characteristics}\label{subsec:res_sfr}
The most important effect of the SP-model feedback can be observed in the star formation history. Fig.~\ref{fig:SFR_isolated} shows the SFR for our isolated galaxy using all three SN feedback schemes considered. The SFR histories for the three feedback schemes differ more significantly during the early phases of the simulations, when the presence of feedback in kinetic form suppress SFR more efficiently. The SP-model feedback, which most effectively suppresses early star formation, has the highest rate of late-time star formation, as material partially ejected by a central ``fountain'' returns at later times to be incorporated in late forming stars. Under NO feedback, the current SFR is 0.23 M$_{\sun}$ yr$^{-1}$, compared to a value of 0.94 under Hot bubble feedback and 1.15 M$_{\sun}$ yr$^{-1}$ under SP-model feedback. The estimated value for the Milky Way is 0.68--1.45 M$_{\sun}$ yr$^{-1}$ \citep{Robitaille2010}. Had we allowed for growing torques from non-axisymmetric tidal forces, these effects would have been more pronounced \citep[see e.g.,][]{Brook14, Uebler14}.

The differences in star formation affect several other galactic characteristics. We therefore measure other important galactic quantities to better understand the impact of each feedback scheme. These measurements are shown in Table \ref{table:results}.

The strongly suppressed early star formation results in an overall younger stellar population, as indicated by the age at which half the new galactic stellar mass is assembled: 1.48 Gyr for SP-model feedback as opposed to 0.37 Gyr for NO feedback and 0.63 for Hot bubble feedback. These half ages are indicated by vertical arrows in Fig.~\ref{fig:SFR_isolated}.

The gas outflow rate (OFR) ---measured as the cumulative outflow rate of gas beyond $\pm$2 kpc from the plane of the disk over 6 Gyr--- suggests the existence of the strong fountain effect in SP-model: OFR is 0.01 M$_{\sun}$ yr$^{-1}$ under NO feedback, 0.37 M$_{\sun}$ yr$^{-1}$ under Hot bubble feedback, and 0.76 M$_{\sun}$ yr$^{-1}$ under SP-model feedback.

The mass loading factor $\eta$ ---measured as the OFR divided by the cumulative stellar mass produced over 6 Gyr--- is a very important metric for assessing the applicability of the models. For the NO feedback case it is 0.01, for the Hot bubble feedback case it is 0.36, and for the SP-model feedback case it is 0.72. \citet{Muratov15} found in their study of galactic winds using high-resolution cosmological simulations ---part of the \textsc{Fire} project--- that galaxies similar in size to our fiducial exhibit gusty outflows at high redshift ($z>1.0$) but then reach a moderate, steady SFR and subdued, weak outflows at low redshift ($z<0.5$). This translates into $\eta$ values close to 10 at high redshift and lower than 1 at low redshift\footnote{\citet{Muratov15} presents several different methods to measure $\eta$ for their simulated galaxies. Our method is most similar to their \textit{Cross T} method, albeit with different geometry in the threshold boundary to categorize the outflow.}.

The different SFR histories of the three feedback scenarios result in different stellar populations after 6 Gyr. Table~\ref{table:results} includes final halo and galactic characteristics. We define the virial radius to be $r_{200}$, the radius where the mean density drops below 200 times the critical density of the universe. We then define the galactic radius to be $r_{10}=0.1r_{200}$. The variations across the feedback scenarios of the virial mass and virial radius are small, as expected. The distinguishable impact of the SP-model lies in the age and mass distribution of the stellar population. 

\begin{figure}
\includegraphics{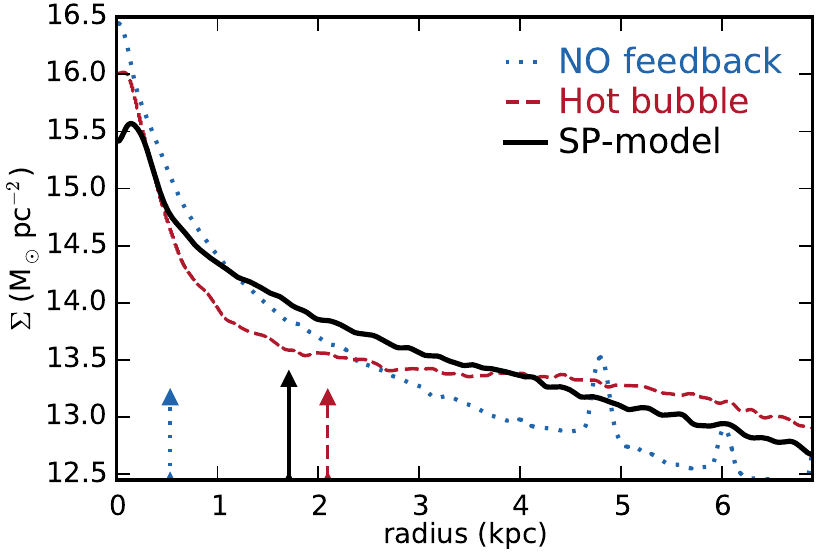}
\caption{Current surface density curves for our simulated galaxy using three different SN feedback schemes. The vertical arrows indicate the effective radius $r_{\mathrm{eff}}$ of a sphere circumscribing half of the new stellar mass.}
\label{fig:sigmacurves}
\end{figure}

The SP-model feedback generates a flatter disk structure in the inner 4 kpc of the galaxy than NO feedback and Hot bubble feedback. Fig.~\ref{fig:sigmacurves} shows surface density ---using dm/dr = 2$\pi r\Sigma(r)$--- at different galactic radii at the end of the simulations for all three SN feedback scenarios. Another indicator of the spread of stellar content is the effective radius $r_{\mathrm{eff}}$ ---the radius of a sphere circumscribing half of the stellar mass (vertical arrows in Fig.~\ref{fig:sigmacurves}). $r_{\mathrm{eff}}$ is larger under SP-model feedback NO feedback by a factor of 3.2, but it is smaller than under Hot bubble feedback by a factor of 0.8. We found no bar formation in our simulations.

\begin{table*}
\centering
\caption{Simulations Results.}
\label{table:results}
\begin{threeparttable}
\begin{tabular}{@{\ \ }l@{\ \ \ }c@{ \ }c@{\ \ }c@{\ \ \ \ \ }c@{ \ }c@{ \ }c@{ \ }c@{\ \ }c@{\ \ \ \ \ }c@{ \ }c@{ \ }c@{ \ }c@{\ \ \ }c@{\ \ \ }c@{\ \ }}
\hline\hline
 & \multicolumn{2}{c}{Halo} & & \multicolumn{4}{c}{Galaxy} & & \multicolumn{6}{c}{Galactic Properties}\\
 & \multicolumn{2}{c}{\hrulefill} & & \multicolumn{4}{c}{\hrulefill} & & \multicolumn{6}{c}{\hrulefill} \\
SN Feedback &
$r_{\mathrm{200}}$ \tnote {a} &
$m_{\mathrm{200}}$ \tnote{b} & &
$r_{\mathrm{eff}}$ \tnote{c} &
$m_{\mathrm{10}}$ \tnote{d} &
$m_{\mathrm{*10}}$ \tnote{e} &
$m_{\mathrm{g10}}$ \tnote{f} & &
$f_{\mathrm{*}}$ \tnote{g} &
$L_X$ \tnote{h} &
$t_{1/2}$ \tnote{i} &
SFR \tnote{j} &
OFR \tnote{k} &
$\eta$ \tnote{l} \\

\multicolumn{1}{c}{Type} & (kpc) & & & (kpc) & & & & &(\%) & (erg s$^{-1}$)  & (Gyr) & \multicolumn{2}{c}{(M$_{\sun}$ yr$^{-1}$)} & \\[0.05 in]
\hline
\\[-1.5mm]
NO feedback & 162.14 & 94.89 & & 0.53 & 18.88 & 0.75 & 0.03 & & 4.7 & 5.1$\times 10^{38}$ & 0.37 & 0.23 & 0.01 & 0.01\\
Hot bubble  & 162.12 & 94.85 & & 2.09 & 18.93 & 0.61 & 0.16 & & 3.8 & 5.8$\times 10^{38}$ & 0.63 & 0.94 & 0.37 & 0.36\\
SP-model    & 162.09 & 94.80 & & 1.71 & 18.86 & 0.63 & 0.10 & & 3.9 & 9.9$\times 10^{38}$ & 1.48 & 1.15 & 0.76 & 0.72\\[2mm]
\hline\hline
\end{tabular}
\begin{tablenotes}
NOTE. -- All masses in units of $10^{10}$ M$_{\sun}$.
\item[a] virial radius; \item[b] virial mass; \item[c] radius of a sphere circumscribing half of the stellar mass; \item[d] galactic mass; \item[e] galactic stellar mass; \item[f] galactic gas mass; \item[g] percentage of galactic baryons converted into stars: $m_{*10}$/($m_{200}\times$ universal baryon fraction);  \item[h] current X-ray luminosity; \item[i] age at which half the new galactic stellar mass is assembled; \item[j] current star formation rate; \item[k] cumulative gas outflow rate from galactic disk over 6 Gyr; \item[l] mass loading factor: cumulative gas outflow from galactic disk divided by the cumulative stellar mass produced over 6 Gyr.
\end{tablenotes}
\end{threeparttable}
\end{table*}

\begin{figure}
\includegraphics{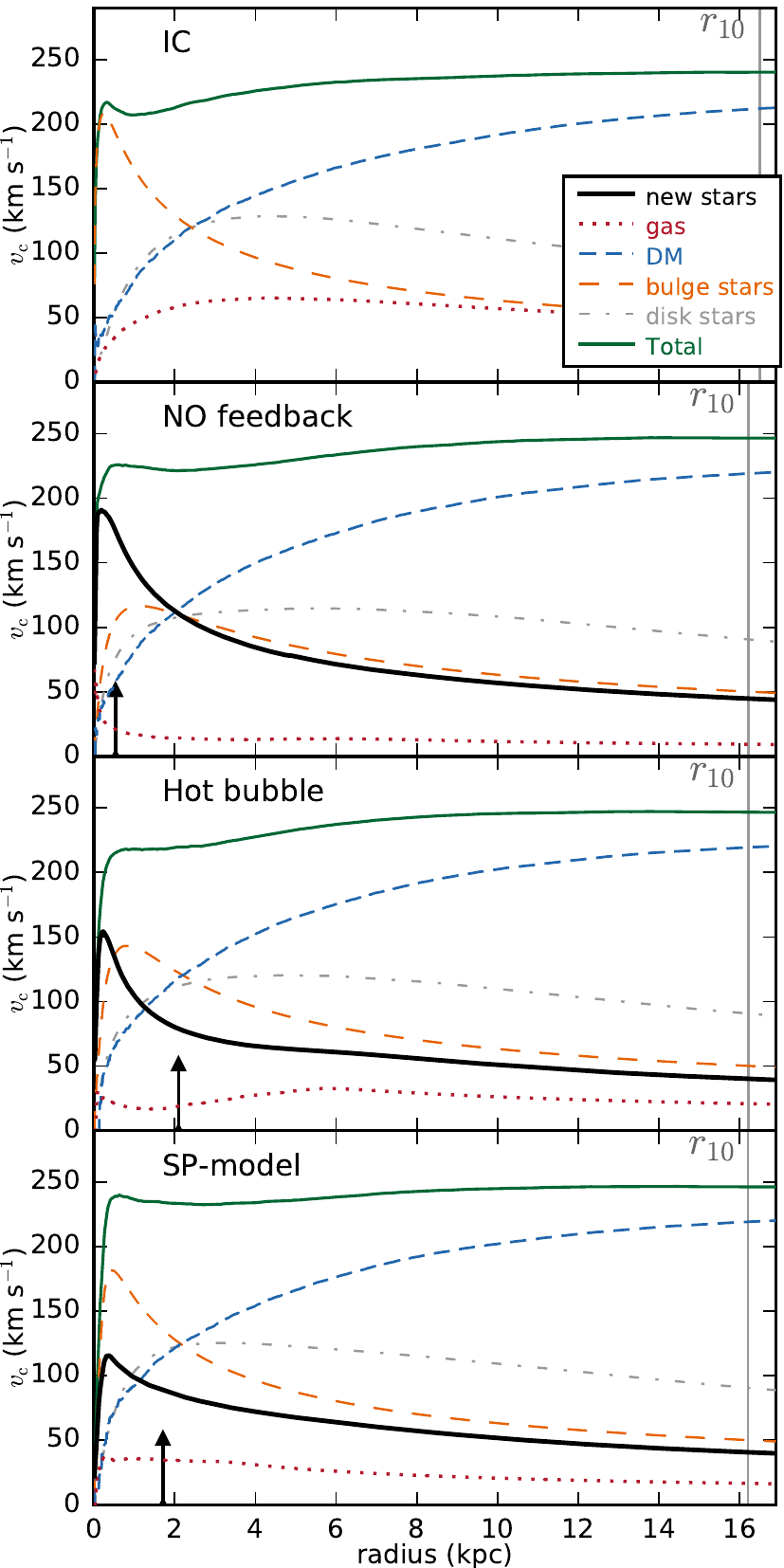}
\caption{Rotational curves for each galactic component for the initial conditions (IC) (upper panel), at end of simulation using a NO feedback scheme (upper middle panel), using a Hot bubble feedback scheme (lower middle panel), and using the SP-model scheme (bottom panel). The galactic radius $r_{10}=0.1 r_{200}$ is indicated with a vertical grey line. The vertical arrows indicate the effective radius $r_{\mathrm{eff}}$ of a sphere circumscribing half of the new stellar mass.}
\label{fig:rot_isolated}
\end{figure}

Fig.~\ref{fig:rot_isolated} shows the theoretical circular velocities $v_{\mathrm{c}}=\sqrt{Gm(r)/r}$ of each galactic component for the initial conditions (IC) (top panel), the NO feedback scenario (upper middle panel), the Hot bubble feedback scenario (lower middle panel), and the SP-model feedback scenario (bottom panel). The SP-model feedback creates a significantly flatter rotational curve for the new stellar component (black solid line) compared to NO feedback and Hot bubble feedback. Under the SP-model feedback, the mean $v_{\mathrm{c}}$ for gas and star particles within $r_{10}$  at the end of the simulation are 24 km s$^{-1}$ and 99 km s$^{-1}$, respectively, compared to 55 km s$^{-1}$ and 69 km s$^{-1}$ in the initial conditions. Interestingly, the old bulge stellar component (orange dashed line) ends up with a significantly flatter rotational curve under the NO feedback and Hot bubble scenarios. We suspect this is caused by a transfer of angular momentum from the bulge star particles to the new star particles, as a significant portion of the star formation occurs near the galactic bulge. The latter is a direct consequence of the absence of energy transfer in the NO feedback scenario and lack of kinetic energy feedback in the Hot bubble scenario, which would otherwise help push out hot gas from star forming regions.

\begin{figure}
\includegraphics{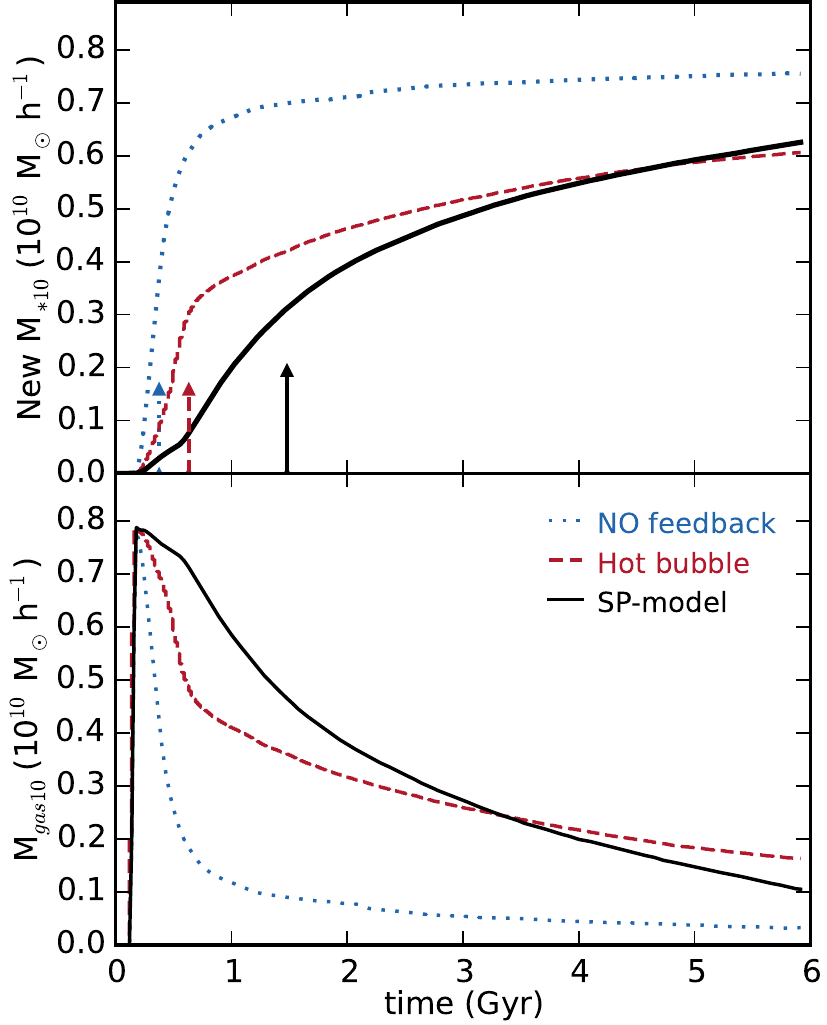}
\caption{New stellar mass (top panel) and gas mass (bottom panel) within the galactic radius $r_{10}=0.1 r_{200}$ as a function of time for three different SN feedback schemes. The vertical arrows indicate the time when the galaxy assembled half of its new stellar mass.}
\label{fig:masses_isolated}
\end{figure}

To compare how stellar populations assemble in the galaxy, Fig.~\ref{fig:masses_isolated} shows a history of stellar (top panel) and gas (bottom panel) masses within the galactic radius $r_{10}$ for all three feedback scenarios. The vertical arrows indicate the time when the galaxy has assembled half of its new stellar mass. In the NO feedback scenario (dotted line), the new stellar content of the galaxy accumulates quickly, whereas any form of energy feedback (Hot bubble or SP-model) hinders this process significantly. Consequently, gas is very quickly depleted under NO feedback, as seen in the bottom panel. Both the Hot bubble and SP-model feedback schemes allow more gas to remain by either heating it or kicking some of it out of the galaxy early on and allowing it to later rain back into the disk. After 6 Gyr the gas-to-stellar-mass ratio is significantly higher when the SP-model is in place (0.16) compared to NO feedback (0.04).

\subsection{Energy Transfer}\label{subsec:res_Etransfer}
\begin{figure}
\includegraphics{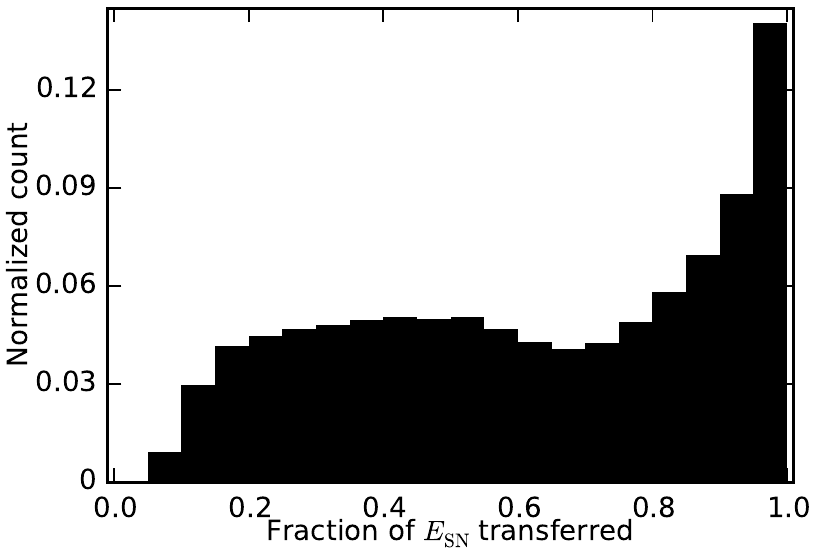}
\caption{Normalized histogram of the fraction of type II SN feedback energy transferred to neighboring gas particles.}
\label{fig:diag_isolated_Etransferred}
\end{figure}

\begin{figure}
\includegraphics{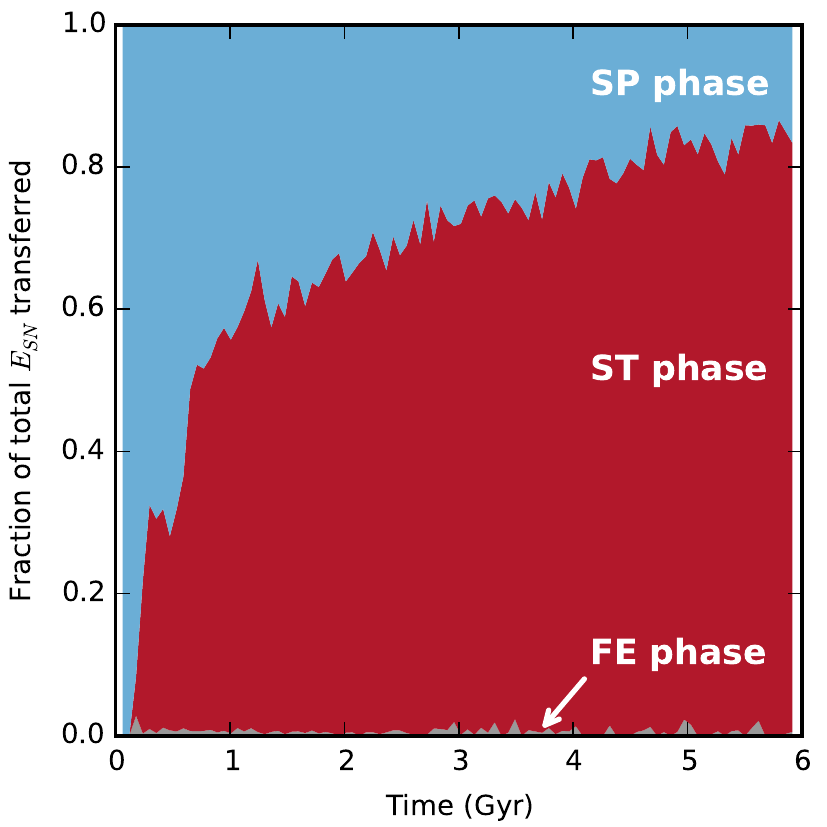}
\caption{Fraction of total transferred type II SN energy at each SNR phase as a function of time when the SP-model energy feedback scheme is implemented on our Milky Way-like isolated galaxy.}
\label{fig:diag_isolated_PhasesE}
\end{figure}

\begin{figure}
\includegraphics{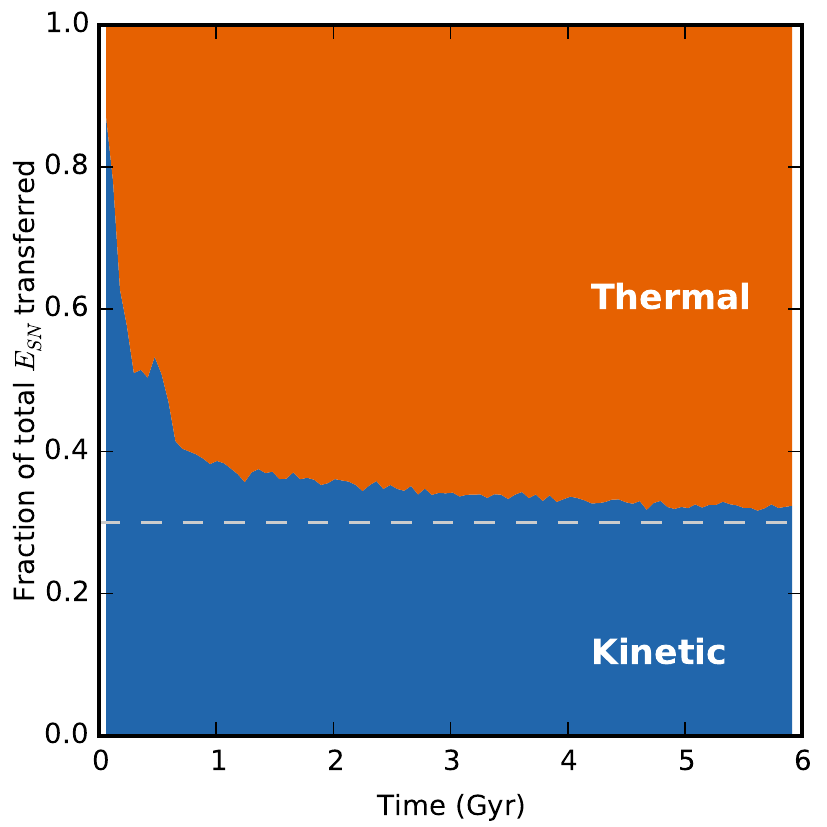}
\caption{Fraction of total transferred type II SN energy as either kinetic (blue) or thermal (orange) form when the SP-model energy feedback scheme is implemented on our Milky Way-like isolated galaxy. The gray dashed line indicates the limit where 30\% of energy is transferred in kinetic form, which is the case during the ST phase of the SP-model feedback scheme.}
\label{fig:diag_isolated_K-Th_sp}
\end{figure}

Figs.~\ref{fig:diag_isolated_Etransferred}--\ref{fig:diag_isolated_K-Th_sp} present a breakdown of the energy transferred in type II SN events. They help to illustrate in more detail how the SP-model feedback distributes energy to the surrounding medium.

According to the description of the SP-model feedback in Sec.~\ref{subsubsec:phases}, some SN energy is radiatively lost during the SP phase (outside $r_{\mathrm{cool}}$), whereas no energy is lost during the FE and ST phases. Therefore, the total amount of energy transferred to gas particles is not always equal to the original SN energy of the ejecta. Fig.~\ref{fig:diag_isolated_Etransferred} shows a normalized histogram of the fraction of type II SN energy released by a star particle that is actually transferred to neighboring gas particles. In $\approx$40\% of the cases, the fraction of the original SN energy transferred is 80\% or greater, either because neighboring gas particles are in FE or ST phase, where no SN energy is lost, or because they are within the SP phase but not too far from the ST--SP transition radius $r_{\mathrm{cool}}$, thus allowing only a small fraction of the original SN ejecta energy to be lost in cooling. By construct, most of the energy in the FE phase, ST phase, and SP phase near $r_{\mathrm{cool}}$ is in thermal form. On the other hand, $\approx$10\% of the cases resulted in an energy transfer less than 20\% of the original SN energy transferred to neighboring gas particles. This is a consequence of gas particles being far into the SP phase, i.e., well outside the ST--SP transition radius $r_{\mathrm{cool}}$, where a significant portion of the original SN energy is lost via radiative cooling. Most of the energy in the SP phase far from $r_{\mathrm{cool}}$ is in kinetic form, as the thermal portion of the remaining energy dissipates more quickly than the kinetic portion (see Eqs.~\ref{E_Th} and \ref{E_K}). Thus, for the cases where most of the original SN energy is dissipated, almost all of the energy transferred is in kinetic form.

Fig.~\ref{fig:diag_isolated_PhasesE} illustrates the fraction of energy transferred at the different SNR phases as a function of time. Most SN energy is transferred to gas particles lying within the ST phase (red region) for three main reasons: first, no energy loss occurs during the ST phase; second, neighboring gas particles in the ST phase have larger SPH kernel weights than those in the SP phase; and third, $r_{\mathrm{cool}}$ increases as ISM densities drop, the latter driven by SN feedback. Also, not many gas particles are found in the FE phase, which is subject to simulation resolution. Finally, Fig.~\ref{fig:diag_isolated_PhasesE} shows that during the first $\approx$1 Gyr of the simulation the majority of the energy is transferred to gas particles in the SP phase. According to Eq.~\ref{STradius}, this indicates that initial high $n_0$ values and low $f_{\mathrm{hot}}$ values --- and thus, small $r_{\mathrm{cool}}$--- must have shifted to lower $n_0$ and higher $f_{\mathrm{hot}}$ values rather rapidly, which in turn increased $r_{\mathrm{cool}}$, such that after $\approx$1 Gyr the energy transferred in the SP phase decreased from almost 100 to $\approx$50\% of the total.

Fig.~\ref{fig:diag_isolated_K-Th_sp} shows the fraction of total energy transferred in the form of kinetic and thermal energy as a function of time. During the first $\approx$1 Gyr, the time during which most energy transfer occurs in the SP phase, the majority of energy transferred is kinetic. This drives early strong winds out of the galactic disk, which help suppress early star formation significantly. Furthermore, the fraction of energy transferred never falls below 30\% (gray dashed line) for 6 Gyr, which illustrates the effect of increasing kinetic energy transfer when the SP phase is included in SN feedback. 

Lastly, we note that in the extreme case of a large hot volume fraction $f_{\mathrm{hot}}$, the amount of transferred thermal and kinetic energy approximates the type of feedback present in the ST phase, not only because the ST--SP transition radius $r_{\mathrm{cool}}$ increases in size (Eq.~\ref{STradius}), but also because the radial factor in Eqs.~\ref{E_Th} and \ref{E_K} approaches unity. We tested a simulated galaxy using exclusively ST feedback on all neighboring gas particles and found that the results were very similar to the results using the SP-model feedback.

\subsection{X-rays}\label{subsec:res_Xrays}
As noted by \citet{Boroson11}, a correlation is expected between the hot gas content and gravitational potential of the galaxy. These authors measured X-ray emission by hot and diffuse gas in galactic halos in the range 0.3--8 keV as proxy for the hot gas content, and central velocity dispersion $\sigma$ as proxy for the gravitational potential, on a sample of 30 early-type galaxies observed with \textit{Chandra}.

\begin{figure}
\includegraphics{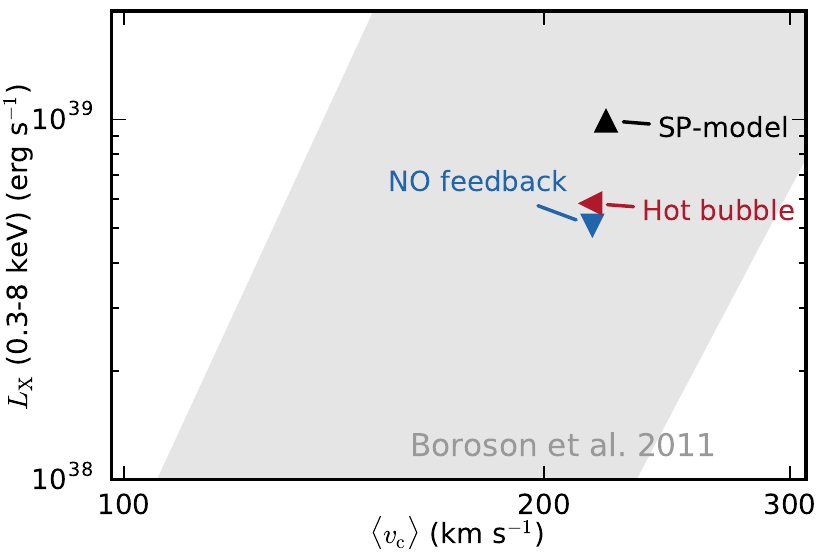}
\caption{X-ray luminosity $L_{\mathrm{X}}$ of our Milky Way-like isolated galaxy versus circular velocity $\langle v_c \rangle$ for three different feedback schemes. We calculate $\langle v_c \rangle$ as the mass-weighted mean rotational velocity for star particles within the galactic radius $r_{10}$. To calculate $L_{\mathrm{X}}$ we use an emissivity prescription that includes bremsstrahlung radiation and line emissions of all tracked species within the 0.3--8 keV band produced by hot ($T \ge 10^6$ K) and diffuse ($\rho \le 2.14\times 10^{-25}$ g cm$^{-3}$) gas halo particles. The observed relation (in light gray) is from \citet{Boroson11}. We translate their measurements of central velocity dispersion $\sigma$ into $\langle v_c \rangle$ using $\langle v_c \rangle^2 \approx 2\sigma^2$.}
\label{fig:Lx_isolated}
\end{figure}

To verify this correlation in our simulated galaxies, we estimate the total galactic X-ray luminosity $L_{\mathrm{X}}$ using an emissivity prescription, described in \citet{Choi14}, that includes bremsstrahlung radiation and line emissions of all tracked species within the 0.3--8 keV band produced by hot ($T \ge 10^6$ K) and diffuse ($\rho \le 2.14\times 10^{-25}$ g cm$^{-3}$) gas halo particles. We then measure circular velocity $\langle v_c \rangle$ as the mass-weighted mean rotational velocity for star particles within the galactic disk $r_{10}$. We compare our $\langle v_c \rangle$ values against the measured $\sigma$ in \citet{Boroson11}. We convert their $\sigma$ values into $\langle v_c \rangle$ using $\langle v_c \rangle^2$ $\approx$ $2\sigma^2$. We find that, given the $\langle v_c \rangle$ values of our simulated galaxies, $L_{\mathrm{X}}$ falls within the observed locus defined in \citet{Boroson11}. Fig.~\ref{fig:Lx_isolated} shows $L_{\mathrm{X}}$ versus $\langle v_c \rangle$ of our isolated galaxy for all three feedback scenarios. Only small $L_{\mathrm{X}}$ differences are seen between the three models, with the SP-model feedback resulting in a $L_{\mathrm{X}}$ value higher than the Hot bubble feedback by a factor of 2. This is probably caused by the large fraction of kinetic energy transferred to the disk gas near SN events at early epochs (see Fig.~\ref{fig:diag_isolated_K-Th_sp}), which pushes it farther into the galaxy halo.

Also, if we consider the mass--$L_{\mathrm{X}}$ relation found by \citet{Anderson15} (see bottom panel of their figure 5) on a large sample of bright galaxies from the Sloan Digital Sky Survey (SDSS), we find that given the current total stellar mass content of our simulated galaxy ($\sim$5$\times 10^{10} M_{\sun}$), its halo $L_{\mathrm{X}}$ falls within the trend observed in the range 10$^{10}-10^{12} M_{\sun}$.

\subsection{Metallicity}\label{subsec:res_metallicity}
Since the different SN feedback mechanisms affect differently the rate and time of star formation, they also affect the final metal content of both gas and star particles in the galaxy. As we described in Sec.~\ref{subsec:sfr}, ejected material from type Ia and II SNe, as well as from AGB and young massive stellar winds, is composed of several metal species in addition to hydrogen and helium.

Under NO feedback, the current median metallicity of the new stellar mass log(Z/Z$_{\sun}$) of our simulated galaxy is 0.11$^{+0.35}_{-0.49}$ (upper and lower bounds indicate the 16th and 84th percentiles); under Hot bubble feedback, it is 0.26$^{+0.33}_{-0.60}$; and under the SP-model feedback, it is 0.34$^{+0.29}_{-0.62}$. The SP-model feedback increases stellar metallicity by a factor of 1.7 compared to NO feedback, and by a factor of 1.2 compared to Hot bubble feedback.

These values, however, do not account for the fact that we do not track metallicity for primordial disk and bulge stars in our galaxy. Therefore, to determine the median metallicity for the entire stellar content, we estimate the median metallicity of this primordial stellar mass using the empirical stellar mass--metallicity relation in \citet{Gallazzi2005}. These authors calculated the median and 16$^{th}$ and 84$^{th}$ percentiles of the distributions in stellar metallicity as a function of stellar mass for a sample of high-quality SDSS spectra of galaxies that includes both early- and late-type galaxies. Given the total mass of the primordial bulge and disk stars in our galaxy (5.3$\times 10^{10}$ M$_{\sun}$), we find that their metallicity ought to be log(Z/Z$_{\sun}$) = $0.04^{+0.22}_{-0.24}$ according to this relation. Using this value, the mass-weighted stellar metallicity for the entire stellar mass content of our galaxy under SP-model feedback is $0.07^{+0.30}_{-0.35}$. This value is statistically similar to the one obtained using the \citet{Gallazzi2005} relation for a galaxy with the same total stellar content as ours. We consider this result only as guidance, since in our isolated galaxy scenario we do not account for external galactic events ---such as mergers or infall of primordial cosmological gas--- that could alter the metal content of galaxies.

\begin{figure}
\includegraphics{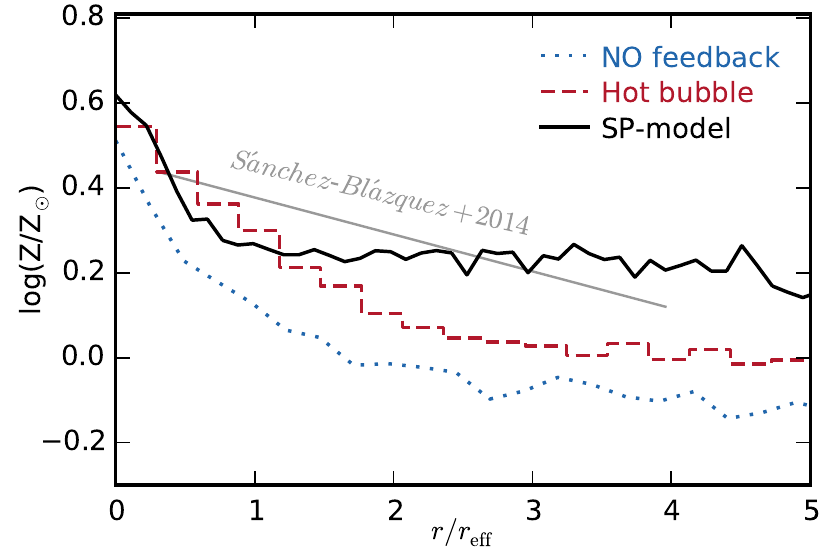}
\caption{Radial distribution of the stellar metallicity content in the galaxy under three different SN feedback schemes. Results from \citet{Sanchez-Blazquez2014} on samples of galaxies from the CALIFA survey are indicated with a gray line. We used the zero point from our own linear regression analysis to indicate their result.}
\label{fig:Metals}
\end{figure}

Fig.~\ref{fig:Metals} shows the stellar radial metallicity distribution in the galaxy under NO feedback (blue dotted line), Hot bubble feedback (red dashed line), and SP-model feedback (solid black line). A linear regression analysis in the inner part of the disk ($r/r_{\mathrm{eff}}<5$) indicates that the stellar metallicity gradient becomes shallower in the SP-model feedback case ($-0.05\pm0.01$) compared to the NO feedback case ($-0.10\pm0.01$) and Hot bubble case ($-0.13\pm0.01$). \citet{Sanchez-Blazquez2014} found for their sample of 62 spiral galaxies from the CALIFA survey a mean slope of $-0.087\pm0.008$. These authors did not specify zero points in their linear fits, and so we used the zero point from our fit to over-plot their result (gray line) in Fig.~\ref{fig:Metals}.

The median galactic gas metallicity increases from 12 + log(O/H) = 8.91$^{+0.14}_{-0.04}$ under NO feedback, to 9.53$\pm0.01$ under Hot bubble feedback, to 9.61$\pm0.02$ under the SP-model feedback. \citet{Tremonti2004} found a median gas metallicity for galaxies similar in stellar content to ours of 12 + log(O/H) = $8.90^{+0.11}_{-0.10}$ in their sample of SDSS galaxies at $0.005 < z < 0.25$. Similarly, \citet{Andrews2013} measure 12 + log(O/H) = 8.75$\pm0.02$ for galaxies similar in stellar content to ours.

The higher-than expected gas metallicity in our simulated galaxy with SP-model feedback may also reflect the inclusion of galaxies at higher redshifts and of early-type galaxies in these empirical samples. Nevertheless, the trend of higher gas metallicity with the SP-model suggests that more metal content is reaching more gas in the disk of the galaxy. The gradient of gas metallicity decreases from $-0.09\pm0.01$ (NO feedback) to $\approx$-0.01 (SP-model feedback). \citet{Sanchez2014} found for their sample of 306 galaxies from the CALIFA survey a characteristic slope of $-0.10\pm0.09$ (with a zero-point of 12+log(O/H) $=8.73\pm0.16$ between 0.3 and 2 $r_{\mathrm{eff}}$).

There is an important effect that may also explain our higher than expected gas metal content. In our own Milky Way galaxy the deuterium abundance is close to 80\% of the primordial abundance, indicating that inflowing pristine and metal-free gas makes up a non negligible fraction of the observed local ISM. Thus, if we made allowance for cosmological infall of gas in our simulations, the metal abundance in our galaxies would be significantly lower.

\subsection{Sensitivity to Ejecta Velocity Parameter}\label{vejecta}
To explore the sensitivity of the SP-model scheme to the ejecta velocity parameter $v_{\mathrm{ej}}$, we re-ran our simulations with the SP-model feedback changing $v_{\mathrm{ej}}$ from our fiducial value of 4500 km s$^{-1}$ to 3000, 6000, and 7500 km s$^{-1}$. We find that the baryon conversion efficiency $f_{*}$ --- defined as $m_{*10}/(m_{200}\times f_{\mathrm{bar}})$, where $f_{\mathrm{bar}}\approx 0.17$--- decreases as $v_{\mathrm{ej}}$ increases: $f_{*}$ goes from 0.042 to 0.032 and 0.024 for $v_{\mathrm{ej}}=3000$, 6000, and 7500 km s$^{-1}$, respectively. Under the NO feedback scenario ---which serves as a proxy for $v_{\mathrm{ej}}=0$ km s$^{-1}$, $f_{*}= 0.047$. The mass loading parameter $\eta$ increases with increasing $v_{\mathrm{ej}}$, from 0.39 to 0.95 and 1.72 for $v_{\mathrm{ej}}=$ 3000, 6000, and 7500 km s$^{-1}$, respectively. Under the NO feedback scenario, $\eta=0.01$. This highlights the sensitivity of the fountain effect to the parameter $v_{\mathrm{ej}}$, particularly when $v_{\mathrm{ej}}$ has a value greater than our fiducial. Lastly, the galactic halo is significantly affected as we increase $v_{\mathrm{ej}}$ beyond our fiducial value: the galactic halo $L_{\mathrm{X}}$ increases by a factor of 11 when we increase $v_{\mathrm{ej}}$ from 4500 to 6000 km s$^{-1}$, and by a factor of 19 when we increase it to 7500 km s$^{-1}$.

There appears to be an energy feedback threshold ---dependent on $v_{\mathrm{ej}}$ according to Eq.~\ref{E_SN}--- beyond which a nascent galaxy at high redshift is strongly compromised by SNR winds. Clearly, this threshold is the point at which the SN ejecta velocity starts reaching the escape velocity of the system. The ejecta velocity would thus set the limit of the halo mass below which the resulting stellar mass would be very small. The peak ratio of stellar to DM mass \citep{Guo10} occurs at stellar mass of 2.2$\times 10^{10}$ M$_{\sun}$, with a strongly observed decline for lower mass systems, which have lower escape velocities.

\subsection{Comparison of Results at Different Resolution Levels}\label{resolutions}
To test convergence of results at different numerical resolution levels, we ran low and high resolution simulations in addition to our fiducial resolution simulation using the SP-model feedback with 10 times lower and 10 times higher mass resolution, respectively. Table~\ref{table:resolutions} describes the SPH particle parameters for the low and high resolution levels.

\begin{table}
\centering
\caption{Comparison of SPH Particles Parameters at Different Resolution Levels.}
\label{table:resolutions}
\begin{threeparttable}
\newcolumntype{d}{D{.}{.}{-1}}
\begin{tabular}{rcdcdcc}
\hline\hline
 & \multicolumn{3}{c}{Low Res} & \multicolumn{3}{c}{High Res} \\
 & \multicolumn{3}{c}{\hrulefill} & \multicolumn{3}{c}{\hrulefill} \\
\multicolumn{1}{c}{Par.} & \multicolumn{1}{c}{N\tnote{a}} & \multicolumn{1}{c}{Mass\tnote{b}} & \multicolumn{1}{c}{$\epsilon$\tnote{c}} & \multicolumn{1}{c}{N\tnote{a}} & \multicolumn{1}{c}{Mass\tnote{b}} & \multicolumn{1}{c}{$\epsilon$\tnote{c}} \\
\multicolumn{1}{c}{Type} & \multicolumn{1}{c}{(10$^4$)} & \multicolumn{1}{c}{(10$^5$)} & \multicolumn{1}{c}{(kpc)} & \multicolumn{1}{c}{(10$^4$)} & \multicolumn{1}{c}{(10$^5$)} & \multicolumn{1}{c}{(kpc)}\\[0.05 in]
\hline
\multicolumn{1}{l}{Gas}   & & & & & & \\
\textit{Halo}  & 0.18 & 13.0 & 0.092 & 17.59 & 0.13 & 0.02 \\
\textit{Disk}  & 0.94 & 13.0 & 0.092 & 94.46 & 0.13 & 0.02 \\
\multicolumn{1}{l}{Stars} & & & & & & \\
\textit{Disk}  & 2.40 & 13.0 & 0.092 & 240.0 & 0.13 & 0.02 \\
\textit{Bulge} & 0.75 & 13.0 & 0.092 & 75.0  & 0.13 & 0.02 \\
\multicolumn{1}{l}{DM} & 3.00 & 446.6 & 0.460 & 300.0 & 4.47 & 0.10 \\[0.05 in]
\hline\hline
\end{tabular}
\begin{tablenotes}
\item[a] Number of particles;
\item[b] per particle, in units of M$_{\sun}$;
\item[c] gravitational softening length.
\end{tablenotes}
\end{threeparttable}
\end{table}

\begin{figure}
\includegraphics{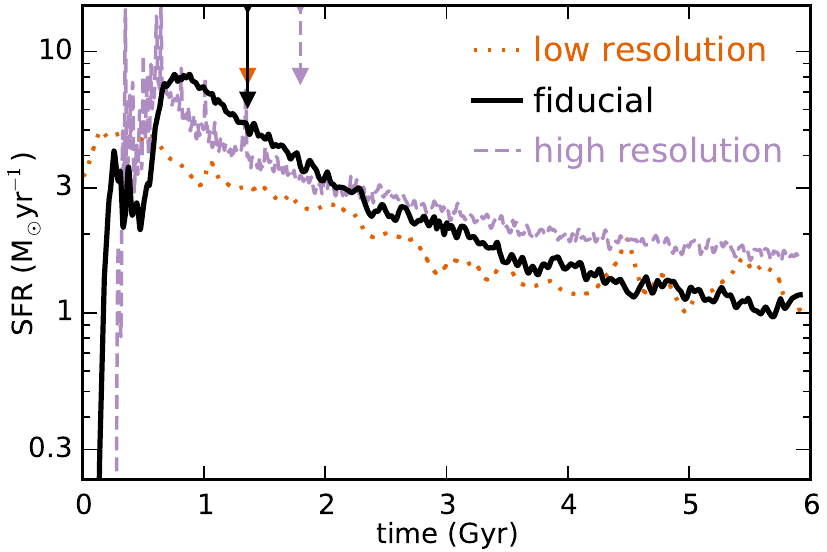}
\caption{SFR as a function of time for our Milky Way-like isolated galaxy using the model SN energy feedback SP-model scheme at low resolution (dotted orange line), our fiducial mid resolution (solid black line), and high resolution (dashed purple line) levels. The vertical arrows indicate the age at which half the new galactic stellar mass has been assembled.}
\label{fig:SFR_res}
\end{figure}

Fig.~\ref{fig:SFR_res} compares the SFR amongst the three resolution levels tested. We find that to first order, SFR converges with resolution, with the high-resolution case having slightly more spikes of star formation at very early epochs than the other two cases. Overall, however, the high resolution case results in higher stellar and gas content in the galaxy: gas-to-star mass ratio goes from 0.3 in the low resolution case to 0.6 in the high resolution case. The high resolution case also results in younger stellar content: $t_{1/2}$ increases from 1.4 Gyr in the low resolution case to 1.8 Gyr in the high resolution case.

\begin{figure}
\includegraphics{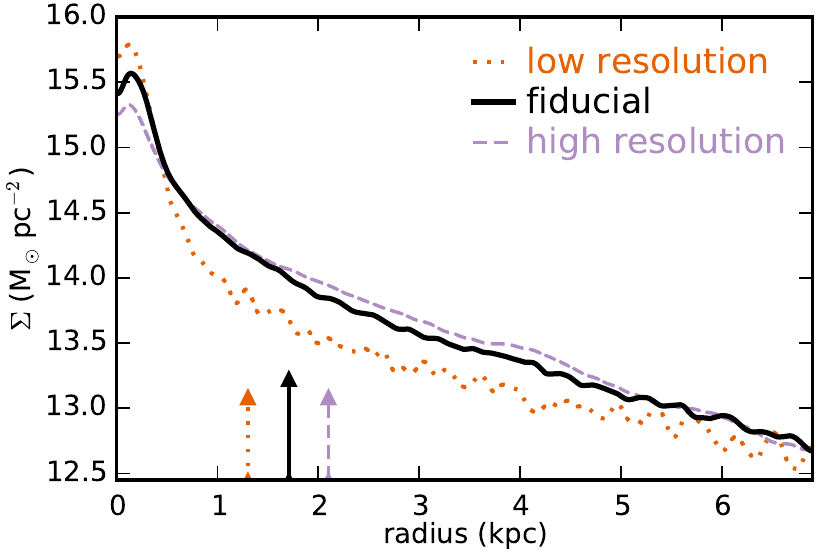}
\caption{Current surface density curves for our simulated galaxy using the model SN energy feedback SP-model scheme at low, mid (fiducial), and high resolution levels. The vertical arrows indicate the effective radius $r_{\mathrm{eff}}$ of a sphere circumscribing half of the new stellar mass.}
\label{fig:sigmacurves_res}
\end{figure}

Similarly, to first order, surface density converges with resolution. However, higher resolution results in a slightly more spread-out, flatter stellar content. Fig.~\ref{fig:sigmacurves_res} shows the surface density curves for the three resolution cases becoming slightly flatter with higher resolution. Also, $r_{\mathrm{eff}}$ (vertical lines in Fig.~\ref{fig:sigmacurves_res}) increases from 1.3 kpc in the low resolution case to 2.1 kpc in the high resolution case. Lastly, the mass loading factor $\eta$ increases to 5.4 in the low resolution case, and to 1.7 in the high resolution case. Several parameter values are compared in Table~\ref{table:results_res} for the three resolution levels.

\begin{table*}
\centering
\caption{Simulations Results at Different Resolution Levels Using the SP-model Feedback.}
\label{table:results_res}
\begin{threeparttable}
\begin{tabular}{@{\ \ }l@{\ \ \ }c@{ \ }c@{\ \ }c@{\ \ \ \ \ }c@{ \ }c@{ \ }c@{ \ }c@{\ \ }c@{\ \ \ \ \ }c@{ \ }c@{ \ }c@{\ \ \ }c@{\ \ \ }c@{\ \ }}
\hline\hline
 & \multicolumn{2}{c}{Halo} & & \multicolumn{4}{c}{Galaxy} & & \multicolumn{5}{c}{Galactic Properties}\\
 & \multicolumn{2}{c}{\hrulefill} & & \multicolumn{4}{c}{\hrulefill} & & \multicolumn{5}{c}{\hrulefill} \\
Resolution &
$r_{\mathrm{200}}$ \tnote {a} &
$m_{\mathrm{200}}$ \tnote{b} & &
$r_{\mathrm{eff}}$ \tnote{c} &
$m_{\mathrm{10}}$ \tnote{d} &
$m_{\mathrm{*10}}$ \tnote{e} &
$m_{\mathrm{g10}}$ \tnote{f} & &
$f_{\mathrm{*}}$ \tnote{g} &
$t_{1/2}$ \tnote{i} &
SFR \tnote{j} &
OFR \tnote{k} &
$\eta$ \tnote{l} \\

\multicolumn{1}{c}{Level} & (kpc) & & & (kpc) & & & & & (\%) & (Gyr) & \multicolumn{2}{c}{(M$_{\sun}$ yr$^{-1}$)} & \\[0.05 in]
\hline
\\[-1.5mm]
Low Res  & 161.53 & 94.13 & & 1.30 & 18.93 & 0.48 & 0.15 & & 3.0 & 1.36 & 1.00 & 4.29 & 5.39\\
Fiducial & 162.09 & 94.80 & & 1.71 & 18.86 & 0.63 & 0.10 & & 3.9 & 1.48 & 1.15 & 0.76 & 0.72\\
High Res & 162.80 & 96.42 & & 2.10 & 19.45 & 0.64 & 0.40 & & 3.9 & 1.79 & 1.62 & 1.78 & 1.67\\[0.05 in]
\hline\hline
\end{tabular}
\begin{tablenotes}
NOTE. -- All masses in units of $10^{10}$ M$_{\sun}$. Notes (a) to (l) are defined in Table~\ref{table:results}.
\end{tablenotes}
\end{threeparttable}
\end{table*}
 
\section{Conclusions}\label{sec:conclusions}
We implement a physically motivated stellar feedback model, including winds from young and old stars, heating from young massive stars, and a SN energy feedback formulation (``SP-model'') that allows for cooling of the original SN ejecta, with the thermal portion dissipating away more quickly than the kinetic portion. The SP-model mimics the known physics that takes place during all phase of SNRs ---free expansion (FE) phase, Sedov-Taylor (ST) phase, and snowplow (SP) phase--- as revealed by both observations and high spatial resolution simulations. It also allows for the important effects of significantly enhanced SN remnant propagation in a multiphase medium. We implement this on an isolated Milky Way-type galaxy with a hot gas halo assembled following the methods in \citet{Springel05a} using the improved TreeSPH code SPHGal \citep{Hu14}, and we evolve it for 6 Gyr.

We also implement wind feedback from both young massive stars and older AGB stars by conserving momentum of the ejected material. For the former, we approximate the ejected material to be commensurate both in mass and velocity with that of type II SN explosions, except that we spread it evenly during the first 3 Myr of the life of the star particle. For the latter, we use mass yield prescriptions found in the literature and an ejecta velocity of 10 km s$^{-1}$, and we implement it throughout several Gyr following type II SN events.

We also implement two mechanisms that limit cooling of the ISM gas, one produced by the Str\"omgren sphere heating surrounding young massive stars, and the other one produced by the radiative recombination processes in dense H\textsc{ii} regions.

In our simulations the SP-model feedback addresses the sub-resolution physics problem in low resolution SPH simulations in which particle distances are barely resolved at the level necessary to allow for FE feedback, the type that many galaxy simulations include. By formulating a type of SN feedback that occurs mainly during the later stages of an SNR, namely ST and SP phases, the SP-model feedback allows for a more accurate representation of the physics involving SN energy feedback.

Given that both ST and SP phases allow for $\geq$ 30\% of the SN ejecta energy to be transferred in kinetic form, and that during SP phase the thermal energy fraction dissipates more quickly with distance than the kinetic fraction, gas particles surrounding SN events receive much stronger momentum kicks while in the ST and SP phases compared to the FE phase, in which feedback is driven by ejecta momentum conservation only. The result of implementing the SP-model feedback is a much flatter SFR history compared to a purely thermal SN feedback (here called Hot bubble feedback) and to having no energy feedback at all. As a consequence, the half age of the formed stars under the SP-model feedback is 4 times later than under no feedback and 2.3 times later than under Hot bubble feedback. The final SFR of the galaxy increases by a factor of 1.2 ---to 1.15 M$_{\sun}$ yr$^{-1}$--- as we go from Hot bubble feedback to the SP-model feedback.

The assembly of the galactic stellar content is more spread out in time in a galaxy with SP-model feedback, as much of the gas kicked out by the SNRs at early times eventually falls back to the galaxy, the so-called galactic ``fountain'' effect, where it cools off and is subject to star formation. The SP-model feedback results in a gas outflow rate from the galaxy of 0.76 M$_{\sun}$ yr$^{-1}$, compared to 0.01 M$_{\sun}$ yr$^{-1}$ under no stellar feedback and 0.37 under Hot bubble feedback. This fountain created by the SP-model feedback helps alleviate the issue of over clustering of stellar content at the galactic center, resulting in flatter rotational curves for the stellar component of the galaxy, flatter surface density curves for the galactic disk, and an overall larger galactic size, with the effective radius increasing by a factor of 3.2 as we go from having no stellar feedback to the SP-model feedback. Furthermore, the SP-model feedback results in a mass loading parameter $\eta$ of 0.72, versus 0.36 under Hot bubble feedback and 0.01 under no feedback.

The integrated stellar metallicity of our galaxy increases by a factor of 1.7 [measured as log(Z/Z$_{\sun}$)], and gas metallicity, by a factor of 5.0 [measured as 12+log(O/H)] as we go from having no stellar feedback to the SP-model feedback. The stellar metallicity disk gradients turn shallower under SP-model feedback ($-0.06$) compared to Hot bubble feedback ($-0.13$).

Given that the parameters of the SP-model ---phase transition radii, energy and mass input--- all scale with the SPH particle masses, our model is only weakly dependent on simulation resolution. The SP-model, however, does not completely address the problem of SN events occurring in overly dense regions in these simulations, which are a pure artificial construct of the limitation in spatial resolution. The unphysical result is SN events occurring in very dense regions that would otherwise be hollowed out by previous SN explosions.

In fact, nature has provided the solution: spatially displacing a significant fraction of type II SN events from their original location would lead to SN energy being released in less dense regions of the ISM. Many OB stars are ``runaways'' ---massive stars with high velocity dispersions resulting from the acceleration given by their SN-exploding, close binary companions. As \citet{Li15} points out, close to 40\% of type II SN explosions are from runaway OB stars, with their high velocity dispersions allowing them to travel up to 100--500 pc from their place of origin. A significant fraction of Type II SN events would, therefore, occur in low-density environments such as above the star forming disk or in inter-spiral arm regions.   

\section*{ACKNOWLEDGMENTS}
We thank Miao Li for valuable comments and feedback and Michaela Hirschmann for her insights into observed galactic metallicities. The simulations used here were run on Columbia University's Yeti cluster and the Max Planck Computing and Data Facility.

\bibliography{references}

\begin{thebibliography}{96}
\expandafter\ifx\csname natexlab\endcsname\relax\def\natexlab#1{#1}\fi

\bibitem[{{Agertz} \& {Kravtsov}(2016)}]{Agertz2016}
{Agertz}, O., \& {Kravtsov}, A.~V. 2016, \apj, 824, 79

\bibitem[{{Agertz} {et~al.}(2013){Agertz}, {Kravtsov}, {Leitner}, \&
  {Gnedin}}]{Agertz13}
{Agertz}, O., {Kravtsov}, A.~V., {Leitner}, S.~N., \& {Gnedin}, N.~Y. 2013,
  ApJ, 770, 25

\bibitem[{{Anderson} {et~al.}(2013){Anderson}, {Bregman}, \&
  {Dai}}]{Anderson13}
{Anderson}, M.~E., {Bregman}, J.~N., \& {Dai}, X. 2013, \apj, 762, 106

\bibitem[{{Anderson} {et~al.}(2015){Anderson}, {Gaspari}, {White}, {Wang}, \&
  {Dai}}]{Anderson15}
{Anderson}, M.~E., {Gaspari}, M., {White}, S.~D.~M., {Wang}, W., \& {Dai}, X.
  2015, MNRAS, 449, 3806

\bibitem[{{Andrews} \& {Martini}(2013)}]{Andrews2013}
{Andrews}, B.~H., \& {Martini}, P. 2013, ApJ, 765, 140

\bibitem[{{Aumer} {et~al.}(2013){Aumer}, {White}, {Naab}, \&
  {Scannapieco}}]{Aumer13}
{Aumer}, M., {White}, S.~D.~M., {Naab}, T., \& {Scannapieco}, C. 2013, \mnras,
  434, 3142

\bibitem[{{Bandiera}(1984)}]{Bandiera84}
{Bandiera}, R. 1984, \aap, 139, 368

\bibitem[{{Bertschinger}(1998)}]{Bertschinger98}
{Bertschinger}, E. 1998, \araa, 36, 599

\bibitem[{{Blondin} {et~al.}(1998){Blondin}, {Wright}, {Borkowski}, \&
  {Reynolds}}]{Blondin98}
{Blondin}, J.~M., {Wright}, E.~B., {Borkowski}, K.~J., \& {Reynolds}, S.~P.
  1998, \apj, 500, 342

\bibitem[{{Boroson} {et~al.}(2011){Boroson}, {Kim}, \& {Fabbiano}}]{Boroson11}
{Boroson}, B., {Kim}, D.-W., \& {Fabbiano}, G. 2011, \apj, 729, 12

\bibitem[{{Brook} {et~al.}(2014){Brook}, {Stinson}, {Gibson}, {Shen},
  {Macci{\`o}}, {Obreja}, {Wadsley}, \& {Quinn}}]{Brook14}
{Brook}, C.~B., {Stinson}, G., {Gibson}, B.~K., {Shen}, S., {Macci{\`o}},
  A.~V., {Obreja}, A., {Wadsley}, J., \& {Quinn}, T. 2014, MNRAS, 443, 3809

\bibitem[{{Chevalier}(1974)}]{Chevalier74}
{Chevalier}, R.~A. 1974, \apj, 188, 501

\bibitem[{{Choi} {et~al.}(2014){Choi}, {Naab}, {Ostriker}, {Johansson}, \&
  {Moster}}]{Choi14}
{Choi}, E., {Naab}, T., {Ostriker}, J.~P., {Johansson}, P.~H., \& {Moster},
  B.~P. 2014, \mnras, 442, 440

\bibitem[{{Choi} {et~al.}(2016){Choi}, {Ostriker}, {Naab}, {Somerville},
  {Hirschmann}, {N{\'u}{\~n}ez}, {Hu}, \& {Oser}}]{Choi2016}
{Choi}, E., {Ostriker}, J.~P., {Naab}, T., {Somerville}, R.~S., {Hirschmann},
  M., {N{\'u}{\~n}ez}, A., {Hu}, C.-Y., \& {Oser}, L. 2016, ArXiv e-prints

\bibitem[{{Cioffi} {et~al.}(1988){Cioffi}, {McKee}, \&
  {Bertschinger}}]{Cioffi88}
{Cioffi}, D.~F., {McKee}, C.~F., \& {Bertschinger}, E. 1988, ApJ, 334, 252

\bibitem[{{Cox}(1972)}]{Cox72}
{Cox}, D.~P. 1972, ApJ, 178, 159

\bibitem[{{Cullen} \& {Dehnen}(2010)}]{Cullen10}
{Cullen}, L., \& {Dehnen}, W. 2010, \mnras, 408, 669

\bibitem[{{Dalla Vecchia} \& {Schaye}(2012)}]{Dalla12}
{Dalla Vecchia}, C., \& {Schaye}, J. 2012, MNRAS, 426, 140

\bibitem[{{Dav{\'e}} {et~al.}(2013){Dav{\'e}}, {Katz}, {Oppenheimer},
  {Kollmeier}, \& {Weinberg}}]{Dave13}
{Dav{\'e}}, R., {Katz}, N., {Oppenheimer}, B.~D., {Kollmeier}, J.~A., \&
  {Weinberg}, D.~H. 2013, MNRAS, 434, 2645

\bibitem[{{Dehnen} \& {Aly}(2012)}]{Dehnen12}
{Dehnen}, W., \& {Aly}, H. 2012, \mnras, 425, 1068

\bibitem[{{Dekel} \& {Silk}(1986)}]{Dekel86}
{Dekel}, A., \& {Silk}, J. 1986, \apj, 303, 39

\bibitem[{Draine(2011)}]{Draine11}
Draine, B. 2011, Physics of the Interstellar and Intergalactic Medium,
  Princeton Series in Astrophysics (Princeton University Press)

\bibitem[{{Dubois} \& {Teyssier}(2008{\natexlab{a}})}]{Dubois08}
{Dubois}, Y., \& {Teyssier}, R. 2008{\natexlab{a}}, \aap, 477, 79

\bibitem[{{Dubois} \& {Teyssier}(2008{\natexlab{b}})}]{Dubois08b}
{Dubois}, Y., \& {Teyssier}, R. 2008{\natexlab{b}}, in Astronomical Society of
  the Pacific Conference Series, Vol. 390, Pathways Through an Eclectic
  Universe, ed. J.~H. {Knapen}, T.~J. {Mahoney}, \& A.~{Vazdekis}, 388

\bibitem[{{Faerman} {et~al.}(2016){Faerman}, {Sternberg}, \&
  {McKee}}]{Faerman2016}
{Faerman}, Y., {Sternberg}, A., \& {McKee}, C.~F. 2016, ArXiv e-prints

\bibitem[{{Furlong} {et~al.}(2015){Furlong}, {Bower}, {Theuns}, {Schaye},
  {Crain}, {Schaller}, {Dalla Vecchia}, {Frenk}, {McCarthy}, {Helly},
  {Jenkins}, \& {Rosas-Guevara}}]{Furlong2015}
{Furlong}, M. {et~al.} 2015, \mnras, 450, 4486

\bibitem[{{Gallazzi} {et~al.}(2005){Gallazzi}, {Charlot}, {Brinchmann},
  {White}, \& {Tremonti}}]{Gallazzi2005}
{Gallazzi}, A., {Charlot}, S., {Brinchmann}, J., {White}, S.~D.~M., \&
  {Tremonti}, C.~A. 2005, MNRAS, 362, 41

\bibitem[{{Girichidis} {et~al.}(2016){Girichidis}, {Walch}, {Naab}, {Gatto},
  {W{\"u}nsch}, {Glover}, {Klessen}, {Clark}, {Peters}, {Derigs}, \&
  {Baczynski}}]{Girichidis16}
{Girichidis}, P. {et~al.} 2016, MNRAS, 456, 3432

\bibitem[{{Guedes} {et~al.}(2011){Guedes}, {Callegari}, {Madau}, \&
  {Mayer}}]{Guedes2011}
{Guedes}, J., {Callegari}, S., {Madau}, P., \& {Mayer}, L. 2011, \apj, 742, 76

\bibitem[{{Guo} {et~al.}(2010){Guo}, {White}, {Li}, \&
  {Boylan-Kolchin}}]{Guo10}
{Guo}, Q., {White}, S., {Li}, C., \& {Boylan-Kolchin}, M. 2010, \mnras, 404,
  1111

\bibitem[{{Haardt} \& {Madau}(2001)}]{Haardt01}
{Haardt}, F., \& {Madau}, P. 2001, in Clusters of Galaxies and the High
  Redshift Universe Observed in X-rays, ed. D.~M. {Neumann} \& J.~T.~V. {Tran}

\bibitem[{{Hernquist}(1990)}]{Hernquist90}
{Hernquist}, L. 1990, \apj, 356, 359

\bibitem[{{Hopkins}(2013)}]{Hopkins2013a}
{Hopkins}, P.~F. 2013, \mnras, 428, 2840

\bibitem[{{Hopkins} {et~al.}(2013){Hopkins}, {Kere{\v s}}, {Murray},
  {Hernquist}, {Narayanan}, \& {Hayward}}]{Hopkins2013b}
{Hopkins}, P.~F., {Kere{\v s}}, D., {Murray}, N., {Hernquist}, L., {Narayanan},
  D., \& {Hayward}, C.~C. 2013, \mnras, 433, 78

\bibitem[{{Hopkins} {et~al.}(2014){Hopkins}, {Kere{\v s}}, {O{\~n}orbe},
  {Faucher-Gigu{\`e}re}, {Quataert}, {Murray}, \& {Bullock}}]{Hopkins14}
{Hopkins}, P.~F., {Kere{\v s}}, D., {O{\~n}orbe}, J., {Faucher-Gigu{\`e}re},
  C.-A., {Quataert}, E., {Murray}, N., \& {Bullock}, J.~S. 2014, MNRAS, 445,
  581

\bibitem[{{Hopkins} {et~al.}(2012){Hopkins}, {Quataert}, \&
  {Murray}}]{Hopkins12}
{Hopkins}, P.~F., {Quataert}, E., \& {Murray}, N. 2012, MNRAS, 421, 3522

\bibitem[{{Hu} {et~al.}(2016){Hu}, {Naab}, {Walch}, {Glover}, \&
  {Clark}}]{Hu15}
{Hu}, C.-Y., {Naab}, T., {Walch}, S., {Glover}, S.~C.~O., \& {Clark}, P.~C.
  2016, \mnras, 458, 3528

\bibitem[{{Hu} {et~al.}(2014){Hu}, {Naab}, {Walch}, {Moster}, \& {Oser}}]{Hu14}
{Hu}, C.-Y., {Naab}, T., {Walch}, S., {Moster}, B.~P., \& {Oser}, L. 2014,
  \mnras, 443, 1173

\bibitem[{{Iwamoto} {et~al.}(1999){Iwamoto}, {Brachwitz}, {Nomoto},
  {Kishimoto}, {Umeda}, {Hix}, \& {Thielemann}}]{Iwamoto99}
{Iwamoto}, K., {Brachwitz}, F., {Nomoto}, K., {Kishimoto}, N., {Umeda}, H.,
  {Hix}, W.~R., \& {Thielemann}, F.-K. 1999, \apjs, 125, 439

\bibitem[{{Janka} {et~al.}(2007){Janka}, {Langanke}, {Marek},
  {Mart{\'{\i}}nez-Pinedo}, \& {M{\"u}ller}}]{Janka07}
{Janka}, H.-T., {Langanke}, K., {Marek}, A., {Mart{\'{\i}}nez-Pinedo}, G., \&
  {M{\"u}ller}, B. 2007, PhysRep, 442, 38

\bibitem[{{Karakas}(2010)}]{Karakas10}
{Karakas}, A.~I. 2010, MNRAS, 403, 1413

\bibitem[{{Katz} {et~al.}(1996){Katz}, {Weinberg}, \& {Hernquist}}]{Katz96}
{Katz}, N., {Weinberg}, D.~H., \& {Hernquist}, L. 1996, \apjs, 105, 19

\bibitem[{{Keller} {et~al.}(2014){Keller}, {Wadsley}, {Benincasa}, \&
  {Couchman}}]{Keller2014}
{Keller}, B.~W., {Wadsley}, J., {Benincasa}, S.~M., \& {Couchman}, H.~M.~P.
  2014, \mnras, 442, 3013

\bibitem[{{Keller} {et~al.}(2015){Keller}, {Wadsley}, \&
  {Couchman}}]{Keller2015}
{Keller}, B.~W., {Wadsley}, J., \& {Couchman}, H.~M.~P. 2015, \mnras, 453, 3499

\bibitem[{{Kere{\v s}} {et~al.}(2009){Kere{\v s}}, {Katz}, {Dav{\'e}},
  {Fardal}, \& {Weinberg}}]{Keres09}
{Kere{\v s}}, D., {Katz}, N., {Dav{\'e}}, R., {Fardal}, M., \& {Weinberg},
  D.~H. 2009, \mnras, 396, 2332

\bibitem[{{Kim} \& {Ostriker}(2015)}]{Kim15}
{Kim}, C.-G., \& {Ostriker}, E.~C. 2015, \apj, 802, 99

\bibitem[{{Kimm} {et~al.}(2015){Kimm}, {Cen}, {Devriendt}, {Dubois}, \&
  {Slyz}}]{Kimm2015}
{Kimm}, T., {Cen}, R., {Devriendt}, J., {Dubois}, Y., \& {Slyz}, A. 2015,
  \mnras, 451, 2900

\bibitem[{{Kudritzki} \& {Puls}(2000)}]{Kudritzki00}
{Kudritzki}, R.-P., \& {Puls}, J. 2000, ARAA, 38, 613

\bibitem[{{Larson}(1974)}]{Larson74}
{Larson}, R.~B. 1974, in IAU Symposium, Vol.~58, The Formation and Dynamics of
  Galaxies, ed. J.~R. {Shakeshaft}, 191

\bibitem[{{Leitherer} {et~al.}(1999){Leitherer}, {Schaerer}, {Goldader},
  {Delgado}, {Robert}, {Kune}, {de Mello}, {Devost}, \&
  {Heckman}}]{Leitherer99}
{Leitherer}, C. {et~al.} 1999, \apjs, 123, 3

\bibitem[{{Li} {et~al.}(2015){Li}, {Ostriker}, {Cen}, {Bryan}, \&
  {Naab}}]{Li15}
{Li}, M., {Ostriker}, J.~P., {Cen}, R., {Bryan}, G.~L., \& {Naab}, T. 2015,
  ApJ, 814, 4

\bibitem[{{Martin}(1999)}]{Martin99}
{Martin}, C.~L. 1999, ApJ, 513, 156

\bibitem[{{Martizzi} {et~al.}(2015){Martizzi}, {Faucher-Gigu{\`e}re}, \&
  {Quataert}}]{Martizzi15}
{Martizzi}, D., {Faucher-Gigu{\`e}re}, C.-A., \& {Quataert}, E. 2015, MNRAS,
  450, 504

\bibitem[{{Monaghan}(1992)}]{Monaghan92}
{Monaghan}, J.~J. 1992, \araa, 30, 543

\bibitem[{{Moster} {et~al.}(2011){Moster}, {Macci{\`o}}, {Somerville}, {Naab},
  \& {Cox}}]{Moster11}
{Moster}, B.~P., {Macci{\`o}}, A.~V., {Somerville}, R.~S., {Naab}, T., \&
  {Cox}, T.~J. 2011, \mnras, 415, 3750

\bibitem[{{Moster} {et~al.}(2013){Moster}, {Naab}, \& {White}}]{Moster13}
{Moster}, B.~P., {Naab}, T., \& {White}, S.~D.~M. 2013, \mnras, 428, 3121

\bibitem[{{Muratov} {et~al.}(2015){Muratov}, {Kere{\v s}},
  {Faucher-Gigu{\`e}re}, {Hopkins}, {Quataert}, \& {Murray}}]{Muratov15}
{Muratov}, A.~L., {Kere{\v s}}, D., {Faucher-Gigu{\`e}re}, C.-A., {Hopkins},
  P.~F., {Quataert}, E., \& {Murray}, N. 2015, MNRAS, 454, 2691

\bibitem[{{Navarro} {et~al.}(1997){Navarro}, {Frenk}, \& {White}}]{Navarro97}
{Navarro}, J.~F., {Frenk}, C.~S., \& {White}, S.~D.~M. 1997, \apj, 490, 493

\bibitem[{{Navarro} \& {White}(1993)}]{Navarro93}
{Navarro}, J.~F., \& {White}, S.~D.~M. 1993, \mnras, 265, 271

\bibitem[{{Newman} {et~al.}(2012){Newman}, {Genzel}, {F{\"o}rster-Schreiber},
  {Shapiro Griffin}, {Mancini}, {Lilly}, {Renzini}, {Bouch{\'e}}, {Burkert},
  {Buschkamp}, {Carollo}, {Cresci}, {Davies}, {Eisenhauer}, {Genel}, {Hicks},
  {Kurk}, {Lutz}, {Naab}, {Peng}, {Sternberg}, {Tacconi}, {Vergani}, {Wuyts},
  \& {Zamorani}}]{Newman12}
{Newman}, S.~F. {et~al.} 2012, \apj, 761, 43

\bibitem[{{Oppenheimer} \& {Dav{\'e}}(2008)}]{Oppenheimer08}
{Oppenheimer}, B.~D., \& {Dav{\'e}}, R. 2008, MNRAS, 387, 577

\bibitem[{{Ostriker} \& {McKee}(1988)}]{Ostriker88}
{Ostriker}, J.~P., \& {McKee}, C.~F. 1988, Reviews of Modern Physics, 60, 1

\bibitem[{{Petruk}(2006)}]{Petruk06}
{Petruk}, O. 2006, ArXiv Astrophysics e-prints

\bibitem[{{Read} \& {Hayfield}(2012)}]{Read12}
{Read}, J.~I., \& {Hayfield}, T. 2012, \mnras, 422, 3037

\bibitem[{{Renaud} {et~al.}(2013){Renaud}, {Bournaud}, {Emsellem}, {Elmegreen},
  {Teyssier}, {Alves}, {Chapon}, {Combes}, {Dekel}, {Gabor}, {Hennebelle}, \&
  {Kraljic}}]{Renaud13}
{Renaud}, F. {et~al.} 2013, MNRAS, 436, 1836

\bibitem[{{Ritchie} \& {Thomas}(2001)}]{Ritchie01}
{Ritchie}, B.~W., \& {Thomas}, P.~A. 2001, \mnras, 323, 743

\bibitem[{{Robitaille} \& {Whitney}(2010)}]{Robitaille2010}
{Robitaille}, T.~P., \& {Whitney}, B.~A. 2010, \apjl, 710, L11

\bibitem[{{Rupke} {et~al.}(2005){Rupke}, {Veilleux}, \& {Sanders}}]{Rupke05}
{Rupke}, D.~S., {Veilleux}, S., \& {Sanders}, D.~B. 2005, \apjs, 160, 115

\bibitem[{{Saitoh} \& {Makino}(2013)}]{Saitoh13}
{Saitoh}, T.~R., \& {Makino}, J. 2013, \apj, 768, 44

\bibitem[{{S{\'a}nchez} {et~al.}(2014){S{\'a}nchez}, {Rosales-Ortega},
  {Iglesias-P{\'a}ramo}, {Moll{\'a}}, {Barrera-Ballesteros}, {Marino},
  {P{\'e}rez}, {S{\'a}nchez-Blazquez}, {Gonz{\'a}lez Delgado}, {Cid Fernandes},
  {de Lorenzo-C{\'a}ceres}, {Mendez-Abreu}, {Galbany}, {Falcon-Barroso},
  {Miralles-Caballero}, {Husemann}, {Garc{\'{\i}}a-Benito}, {Mast}, {Walcher},
  {Gil de Paz}, {Garc{\'{\i}}a-Lorenzo}, {Jungwiert}, {V{\'{\i}}lchez},
  {J{\'{\i}}lkov{\'a}}, {Lyubenova}, {Cortijo-Ferrero}, {D{\'{\i}}az},
  {Wisotzki}, {M{\'a}rquez}, {Bland-Hawthorn}, {Ellis}, {van de Ven}, {Jahnke},
  {Papaderos}, {Gomes}, {Mendoza}, \& {L{\'o}pez-S{\'a}nchez}}]{Sanchez2014}
{S{\'a}nchez}, S.~F. {et~al.} 2014, AAP, 563, A49

\bibitem[{{S{\'a}nchez-Bl{\'a}zquez} {et~al.}(2014){S{\'a}nchez-Bl{\'a}zquez},
  {Rosales-Ortega}, {M{\'e}ndez-Abreu}, {P{\'e}rez}, {S{\'a}nchez}, {Zibetti},
  {Aguerri}, {Bland-Hawthorn}, {Catal{\'a}n-Torrecilla}, {Cid Fernandes}, {de
  Amorim}, {de Lorenzo-Caceres}, {Falc{\'o}n-Barroso}, {Galazzi},
  {Garc{\'{\i}}a Benito}, {Gil de Paz}, {Gonz{\'a}lez Delgado}, {Husemann},
  {Iglesias-P{\'a}ramo}, {Jungwiert}, {Marino}, {M{\'a}rquez}, {Mast},
  {Mendoza}, {Moll{\'a}}, {Papaderos}, {Ruiz-Lara}, {van de Ven}, {Walcher}, \&
  {Wisotzki}}]{Sanchez-Blazquez2014}
{S{\'a}nchez-Bl{\'a}zquez}, P. {et~al.} 2014, AAP, 570, A6

\bibitem[{{Scannapieco} {et~al.}(2006){Scannapieco}, {Tissera}, {White}, \&
  {Springel}}]{Scannapieco06}
{Scannapieco}, C., {Tissera}, P.~B., {White}, S.~D.~M., \& {Springel}, V. 2006,
  \mnras, 371, 1125

\bibitem[{{Schaye} {et~al.}(2015){Schaye}, {Crain}, {Bower}, {Furlong},
  {Schaller}, {Theuns}, {Dalla Vecchia}, {Frenk}, {McCarthy}, {Helly},
  {Jenkins}, {Rosas-Guevara}, {White}, {Baes}, {Booth}, {Camps}, {Navarro},
  {Qu}, {Rahmati}, {Sawala}, {Thomas}, \& {Trayford}}]{Schaye15}
{Schaye}, J. {et~al.} 2015, MNRAS, 446, 521

\bibitem[{Sedov(1959)}]{Sedov59}
Sedov, L. 1959, Similarity and Dimensional Methods in Mechanics (New York:
  Academic Press)

\bibitem[{{Simpson} {et~al.}(2015){Simpson}, {Bryan}, {Hummels}, \&
  {Ostriker}}]{Simpson15}
{Simpson}, C.~M., {Bryan}, G.~L., {Hummels}, C., \& {Ostriker}, J.~P. 2015,
  ApJ, 809, 69

\bibitem[{{Somerville} \& {Dav{\'e}}(2015)}]{Somerville15}
{Somerville}, R.~S., \& {Dav{\'e}}, R. 2015, ARAA, 53, 51

\bibitem[{{Springel}(2005)}]{Springel05b}
{Springel}, V. 2005, \mnras, 364, 1105

\bibitem[{{Springel}(2010)}]{Springel10}
---. 2010, \araa, 48, 391

\bibitem[{{Springel} {et~al.}(2005){Springel}, {Di Matteo}, \&
  {Hernquist}}]{Springel05a}
{Springel}, V., {Di Matteo}, T., \& {Hernquist}, L. 2005, \mnras, 361, 776

\bibitem[{{Springel} \& {Hernquist}(2002)}]{Springel02}
{Springel}, V., \& {Hernquist}, L. 2002, \mnras, 333, 649

\bibitem[{{Springel} \& {Hernquist}(2003)}]{Springel03}
---. 2003, \mnras, 339, 289

\bibitem[{{Steidel} {et~al.}(2010){Steidel}, {Erb}, {Shapley}, {Pettini},
  {Reddy}, {Bogosavljevi{\'c}}, {Rudie}, \& {Rakic}}]{Steidel10}
{Steidel}, C.~C., {Erb}, D.~K., {Shapley}, A.~E., {Pettini}, M., {Reddy}, N.,
  {Bogosavljevi{\'c}}, M., {Rudie}, G.~C., \& {Rakic}, O. 2010, \apj, 717, 289

\bibitem[{{Stinson} {et~al.}(2006){Stinson}, {Seth}, {Katz}, {Wadsley},
  {Governato}, \& {Quinn}}]{Stinson06}
{Stinson}, G., {Seth}, A., {Katz}, N., {Wadsley}, J., {Governato}, F., \&
  {Quinn}, T. 2006, \mnras, 373, 1074

\bibitem[{{Stinson} {et~al.}(2013){Stinson}, {Brook}, {Macci{\`o}}, {Wadsley},
  {Quinn}, \& {Couchman}}]{Stinson13}
{Stinson}, G.~S., {Brook}, C., {Macci{\`o}}, A.~V., {Wadsley}, J., {Quinn},
  T.~R., \& {Couchman}, H.~M.~P. 2013, MNRAS, 428, 129

\bibitem[{{Strickland} {et~al.}(2000){Strickland}, {Heckman}, {Weaver}, \&
  {Dahlem}}]{Strickland00}
{Strickland}, D.~K., {Heckman}, T.~M., {Weaver}, K.~A., \& {Dahlem}, M. 2000,
  \aj, 120, 2965

\bibitem[{{Str{\"o}mgren}(1939)}]{Stromgren39}
{Str{\"o}mgren}, B. 1939, ApJ, 89, 526

\bibitem[{{Su} {et~al.}(2016){Su}, {Hopkins}, {Hayward}, {Faucher-Giguere},
  {Keres}, {Ma}, \& {Robles}}]{Su2016}
{Su}, K.-Y., {Hopkins}, P.~F., {Hayward}, C.~C., {Faucher-Giguere}, C.-A.,
  {Keres}, D., {Ma}, X., \& {Robles}, V.~H. 2016, ArXiv e-prints

\bibitem[{{Taylor}(1950)}]{Taylor50}
{Taylor}, G. 1950, Proceedings of the Royal Society of London Series A, 201,
  159

\bibitem[{{Teyssier} {et~al.}(2013){Teyssier}, {Pontzen}, {Dubois}, \&
  {Read}}]{Teyssier13}
{Teyssier}, R., {Pontzen}, A., {Dubois}, Y., \& {Read}, J.~I. 2013, MNRAS, 429,
  3068

\bibitem[{{Tremonti} {et~al.}(2004){Tremonti}, {Heckman}, {Kauffmann},
  {Brinchmann}, {Charlot}, {White}, {Seibert}, {Peng}, {Schlegel}, {Uomoto},
  {Fukugita}, \& {Brinkmann}}]{Tremonti2004}
{Tremonti}, C.~A. {et~al.} 2004, ApJ, 613, 898

\bibitem[{{{\"U}bler} {et~al.}(2014){{\"U}bler}, {Naab}, {Oser}, {Aumer},
  {Sales}, \& {White}}]{Uebler14}
{{\"U}bler}, H., {Naab}, T., {Oser}, L., {Aumer}, M., {Sales}, L.~V., \&
  {White}, S.~D.~M. 2014, MNRAS, 443, 2092

\bibitem[{{Vogelsberger} {et~al.}(2013){Vogelsberger}, {Genel}, {Sijacki},
  {Torrey}, {Springel}, \& {Hernquist}}]{Vogelsberger13}
{Vogelsberger}, M., {Genel}, S., {Sijacki}, D., {Torrey}, P., {Springel}, V.,
  \& {Hernquist}, L. 2013, MNRAS, 436, 3031

\bibitem[{{Vogelsberger} {et~al.}(2014){Vogelsberger}, {Genel}, {Springel},
  {Torrey}, {Sijacki}, {Xu}, {Snyder}, {Nelson}, \&
  {Hernquist}}]{Vogelsberger14}
{Vogelsberger}, M. {et~al.} 2014, MNRAS, 444, 1518

\bibitem[{{Walch} \& {Naab}(2015)}]{Walch15}
{Walch}, S., \& {Naab}, T. 2015, MNRAS, 451, 2757

\bibitem[{{White} \& {Frenk}(1991)}]{White91}
{White}, S.~D.~M., \& {Frenk}, C.~S. 1991, \apj, 379, 52

\bibitem[{{Woosley} \& {Weaver}(1995)}]{Woosley95}
{Woosley}, S.~E., \& {Weaver}, T.~A. 1995, \apjs, 101, 181

\end{thebibliography}

\appendix

\section{Description of the SP-model for SPH Simulations.}\label{appendix}
As described in Sec.~\ref{subsubsec:implementation}, the SP-model feedback formulation states that a neighboring gas particle \textit{j} can receive energy from the SN event in one of three different ways, depending on its distance to the star particle relative to the phase transition radii $r_{\mathrm{ST}}$ and $r_{\mathrm{cool}}$ (Eqs. \ref{FEradius} and \ref{STradius}).

The FE-ST transition radius $r_{\mathrm{ST}}$ depends on the density of the medium, and the ST-SP transition radius $r_{\mathrm{cool}}$ depends on both the density and the temperature of the medium, the latter via the dependency of the hot volume fraction $f_{\mathrm{hot}}$ on the gas temperature (Eq.~\ref{fhot}). To allow for sharp density and temperature gradients around the star particle, we calculate both $r_{\mathrm{ST}}$ and $r_{\mathrm{cool}}$ for each of the closest 10 neighboring gas particles using the gas particle density $\rho_j$ and temperature $T_j$. The total amount of energy assigned to particle \textit{j} from the SN event is
\begin{equation}\label{delE}
\Delta E_j = w_j\ E_{\mathrm{SN}},
\end{equation}
where $w_j$ is the SPH kernel weight assigned to particle \textit{j} (in our simulations we use the Wendland $C^4$ kernel). The transfer of this energy affects three basic quantities in particle \textit{j}: its mass, its velocity, and its entropy (via addition of thermal energy). For the mass, one simply adds to the mass of particle \textit{j}, $M_j$, the value $\Delta m_j = w_j\ M_{ej}$. How the other two quantities change varies from one SNR phase to the other.

\subsection{FE Phase}
If particle \textit{j} is in the FE phase, then momentum conservation applies. The velocity vector to add to particle \textit{j}, assuming a perfectly non-elastic collision with the ejecta, is
\begin{equation}\label{FE_vadd}
\boldsymbol{\Delta v_{\mathrm{FE}}} = \frac{\epsilon}{1+\epsilon}\ (\boldsymbol{v_{\mathrm{ej}}} -\boldsymbol{v_j}),
\end{equation}
where $\epsilon=\Delta m_j/M_j$, $\boldsymbol{v_{\mathrm{ej}}}$ is the ejecta velocity and $\boldsymbol{v_j}$ is the velocity of particle \textit{j} before the collision. All three velocities are with respect to the star particle's rest frame. $\boldsymbol{\Delta v_{\mathrm{FE}}}$ is added to $\boldsymbol{v_j}$ radially away from the star particle.

To conserve energy, one must increase the thermal energy of particle \textit{j} by
\begin{equation}\label{FE_Ethadd}
\Delta E_{\mathrm{Th,FE}} = \frac{1}{2}\ \mu \ (\boldsymbol{v_{\mathrm{ej}}} -\boldsymbol{v_j})^2,
\end{equation}
where $\mu=\Delta m_j\ M_j/(\Delta m_j+M_j)$.

\subsection{ST Phase}
If particle \textit{j} is in the ST phase, then the analytical self-similar solution applies, which gives a constant breakdown of SN energy between $\sim$30\% kinetic and $\sim$70\% thermal. This means that one must increase the thermal energy of particle \textit{j} by $\Delta E_{\mathrm{Th,ST}} = 0.7\ \Delta E_j$. In general, the change in kinetic energy for particle \textit{j} is $\Delta E_{\mathrm{K}} = E_{\mathrm{K,new}} - E_{\mathrm{K,old}}$, or
\begin{equation}\label{delK2}
\Delta E_{\mathrm{K}} = \frac{1}{2}(M_j+\Delta m_j)(\boldsymbol{v_j} + \boldsymbol{\Delta v})^2 - \frac{1}{2}M_jv_j^2,
\end{equation}
where $\boldsymbol{\Delta v}$ is some velocity added to particle \textit{j}. In the case of ST phase, the velocity vector $\boldsymbol{\Delta v_{\mathrm{ST}}}$ to add to particle \textit{j} is such that
\begin{equation}\label{ST_EKadd}
\Delta E_{\mathrm{K,ST}} = 0.3\ \Delta E_j = \frac{1}{2}\ \Delta m_j\ \Delta v_{\mathrm{ST}}^2.
\end{equation}
Combining Eqs. \ref{delK2} and \ref{ST_EKadd} ---with $\boldsymbol{\Delta v_{\mathrm{ST}}}$ in lieu of $\boldsymbol{\Delta v}$--- and working out the vector algebra, one finds that 
\begin{equation}\label{ST_vadd}
\boldsymbol{\Delta v_{\mathrm{ST}}} = \alpha_{\mathrm{ST}}\ \boldsymbol{v_{\mathrm{ej}}};
\end{equation}
here $\boldsymbol{\Delta v_{\mathrm{ST}}}$ has the same direction as $\boldsymbol{v_{\mathrm{ej}}}$, and $\alpha_{\mathrm{ST}}$ is
\begin{equation}\label{alpha_st}
\alpha_{\mathrm{ST}} = \frac{\sqrt{|0.3\frac{\epsilon}{1+\epsilon}v_{\mathrm{ej}}^2 + (\frac{1}{1+\epsilon}- \mathrm{sin}^2\theta)v_j^2|} - v_j\mathrm{cos}\ \theta}{v_{\mathrm{ej}}},
\end{equation}
where $\theta$ is the angle between $v_{\mathrm{ej}}$ and $v_j$. As in the FE phase, All velocities are with respect to the star particle's rest frame, and $\boldsymbol{\Delta v_{\mathrm{ST}}}$ is added to $\boldsymbol{v_j}$ radially away from the star particle.

\subsection{SP Phase}
If particle \textit{j} is in the SP phase, then Eqs. \ref{E_Th} and \ref{E_K} apply. One must increase the thermal energy of particle \textit{j} by
\begin{equation}\label{SP_Ethadd}
\Delta E_{\mathrm{Th,SP}} = \frac{0.7 \Delta E_j}{1 + 0.5(r_j / r_{\mathrm{cool}})^6},
\end{equation}
where $r_j$ is the distance between the star particle and particle \textit{j}. Similarly, one must increase the kinetic energy of particle \textit{j} by
\begin{equation}\label{SP_EKadd}
\Delta E_{\mathrm{K,SP}} = \frac{0.3 \Delta E_j}{1 + 0.18(r_j / r_{\mathrm{cool}})^3}.
\end{equation}

Mimicking the steps for the ST phase and defining
\begin{equation}\label{facSP}
f_{\mathrm{SP}} = \frac{1}{1 + 0.18(r_j / r_{\mathrm{cool}})^3},
\end{equation}
one combines Eqs. \ref{delE}, \ref{delK2}, \ref{SP_EKadd}, \ref{facSP} and \ref{E_SN} to find that the velocity vector $\boldsymbol{\Delta v_{\mathrm{SP}}}$ to add to particle \textit{j} is
\begin{equation}\label{SP_vadd}
\boldsymbol{\Delta v_{\mathrm{SP}}} = \alpha_{\mathrm{SP}}\ \boldsymbol{v_{\mathrm{ej}}},
\end{equation}
where $\boldsymbol{\Delta v_{\mathrm{SP}}}$ has the same direction as $\boldsymbol{v_{\mathrm{ej}}}$, and $\alpha_{\mathrm{SP}}$ is
\begin{equation}\label{alpha_sp}
\alpha_{\mathrm{SP}} = \frac{\sqrt{|0.3f_{\mathrm{SP}}\frac{\epsilon}{1+\epsilon}v_{\mathrm{ej}}^2 + (\frac{1}{1+\epsilon}-\mathrm{sin}^2\theta)v_j^2|} - v_j\mathrm{cos}\ \theta}{v_{\mathrm{ej}}}.
\end{equation}
Again, all velocities are with respect to the star particle's rest frame, and $\boldsymbol{\Delta v_{\mathrm{SP}}}$ is added to $\boldsymbol{v_j}$ radially away from the star particle.

\end{document}